\documentclass[10pt,journal,twoside]{IEEEtran}
\usepackage{multirow}
\usepackage{amssymb}
\usepackage[dvips]{graphicx}
\usepackage{amsmath}
\usepackage{amsthm}
\usepackage{latexsym,bm}
\usepackage{color}
\usepackage{subfigure}
\usepackage{longtable}
\usepackage{accents}
\usepackage{cite}
\usepackage{enumerate}
\usepackage{arydshln}
\usepackage{slashbox}
\usepackage{algorithmic}
\usepackage{booktabs}
\usepackage{subfigure}
\usepackage{stfloats}
\usepackage[table,xcdraw]{xcolor}
\usepackage{graphicx}
\usepackage{bm}

   \def\bu{{\mathbf{u}}}

  \def\br{{\mathbf{r}}}

\def\bg{{\mathbf{g}}} 

\def\bA{{\mathbf{A}}}   
\def\bB{{\mathbf{B}}}   \def\bV{{\mathbf{V}}}

\def\bE{{\mathbf{E}}}   
\def\bF{{\mathbf{F}}}   

\makeatletter
\def\widebar{\accentset{{\cc@style\underline{\mskip10mu}}}}
\def\Widebar{\accentset{{\cc@style\underline{\mskip8mu}}}}
\makeatother

\theoremstyle{plain}

\theoremstyle{definition}
\theoremstyle{definition}

\begin{document}
\title{A New Frequency-Bin-Index LoRa System for High-Data-Rate Transmission: Design and Performance Analysis
\thanks{H.~Ma, Y.~Fang, G.~Cai, and G.~Han are with the School of Information Engineering, Guangdong University of Technology, Guangzhou 510006, China, (email: mh-zs@163.com; \{fangyi,~caiguofa2006,~gjhan\}@gdut.edu.cn).}
\thanks{Y.~Li is with the School of Electrical and Information Engineering, The University of Sydney, Sydney, NSW 2006, Australia (e-mail: yonghui.li@sydney.edu.au).}}
\author{Huan Ma, Yi Fang, {\em Member, IEEE}, Guofa Cai, {\em Member, IEEE}, \\Guojun Han, {\em Senior Member, IEEE}, Yonghui Li, {\em Fellow, IEEE}}
\maketitle
\begin{abstract}
As an attempt to tackle the low-data-rate issue of the conventional LoRa systems, we propose two novel frequency-bin-index (FBI) LoRa schemes. In scheme I, the indices of starting frequency bins (SFBs) are utilized to carry the information bits. To facilitate the actual implementation, the SFBs of each LoRa signal are divided into several groups prior to the modulation process in the proposed FBI-LoRa system.
To further improve the system flexibility, we formulate a generalized modulation scheme and propose scheme II by treating the SFB groups as an additional type of transmission entity. In scheme II, the combination of SFB indices and that of SFB group indices are both exploited to carry the information bits.
We derive the theoretical expressions for bit-error-rate (BER) and throughput of the proposed FBI-LoRa system with two modulation schemes over additive white Gaussian noise (AWGN) and Rayleigh fading channels.
Theoretical and simulation results show that the proposed FBI-LoRa schemes can significantly increases the transmission throughput compared with the existing LoRa systems at the expense of a slight loss in BER performance. Thanks to the appealing superiorities, the proposed FBI-LoRa system is a promising alternative for high-data-rate Internet of Things (IoT) applications.

\end{abstract}

\section{Introduction}
Internet of Things (IoT) technologies are progressively playing an important role in our life \cite{9257428}. According to the Ericsson Mobile Report, it is predicted that by 2022, approximately 29 billion IoT devices will be connected via wireless technology \cite{121213567}. Driven by the massive connectivity and low-power consumption requirements in IoT applications, low-power wide area (LPWA) network has emerged as a promising solution and has been widely deployed in many fields \cite{9269490,9207749}. Among all the prospective LPWA technologies, LoRa that operates in unlicensed frequency bands has attracted much attention \cite{9093735,8626078}.

LoRa is a chirp spread-spectrum (CSS) based modulation with low power, low data rate and long range \cite{8959186,8883217,9352969}. In the LoRa modulation, a multidimensional space for LoRa signals is formed by cyclic shifts of chirp signal with linearly increased frequency, and different LoRa signals are mutually orthogonal \cite{9018296,9000860,8067462,9207749}. The frequency of these LoRa chirp signals linearly increase at different rate, which depends on the spreading factor ($SF=7,\ldots,12$) \cite{8392707,8903531}. Increasing the spreading factor can expand the LoRa coverage area, but reduce the data rate simultaneously. Under the leadership of the LoRa Alliance,\footnote{https://lora-alliance.org} LoRa has gained more and more commercial traction, and hence has been widely used in various fields, such as manufacturing\cite{8390453,9294099}, logistics industry\cite{SINHA201714}, infrastructure \cite{ZORBAS20201,9055222}, and smart home \cite{9065199}. With the notable success of LoRa in industry, a significant amount of research attention has been paid to LoRa. The initial works have focused on experimental investigation, which evaluate various performance indicators of LoRa modulation, such as coverage capability \cite{7803607} and sensitivity \cite{8268120}.
In particular, the scalability \cite{8090518} and capacity \cite{8839056} of LoRa have been extensively studied, where ALOHA network is considered to characterize LoRa-based LPWA network \cite{9018210}. With the implementation of reverse engineering for the LoRa physical layer \cite{77566824}, corresponding theoretical research has become active in recent years. Based on the rigorous mathematical description of LoRa modulation \cite{8067462}, the waveform and spectral characteristics of LoRa modulation have been studied in \cite{8723130}. The theoretical error performance of LoRa modulation in different scenarios has been analyzed in \cite{8392707,sensosloraber}.

Although LoRa modulation can adjust the data rate by varying the spreading factor, its maximum rate is still too low to satisfy the requirement of many potential applications, such as smart home/building\cite{9151815,8710297}, image transmission\cite{9177489}, and indoor IoT\cite{8827665}. To overcome this weakness, some works have modified the conventional LoRa modulation to increase the maximum data rate, such as interleaved chirp spreading (ICS) LoRa\cite{8607020}, slope-shift keying (SSK) LoRa\cite{9123393}, and phase shift keying (PSK) LoRa\cite{8746470}. Specifically, both ICS-LoRa and SSK-LoRa obtain a new signal dimension by modifying the conventional LoRa signal, which can be utilized to carry more information. PSK-LoRa divides the information bits into two groups, where the first group is utilized to select the starting frequency bin (SFB) of the LoRa signal, and the second group is used for PSK modulation to determine the initial phase of the LoRa signal. Although the above methods can increase the data rate of LoRa modulation, but the improvement is marginal. In \cite{7815384}, the authors have pointed out that it is necessary to further optimize the LoRa modulations in order to satisfy the data-rate requirements of different application scenarios.

Index modulation (IM) is another desirable solution to increase the data rate of LoRa modulation. IM is an emerging type of signal modulation that carries the signal through the index(es) of certain transmission entities. These transmission entities can be either practical (e.g., antennas, frequency carriers, and subcarriers) or virtual (e.g., time slots, space-time matrix, and antenna activation sequence) \cite{IM2017,9233391}.
As a typical instance, in the well-known orthogonal frequency division multiplexing (OFDM) IM system \cite{6587554}, the transmitter not only uses
the $M$-ary signal constellation to carry information bits, but also exploits the index combination of activated subcarriers to
carry additional information bits. On the basis of OFDM-IM, researchers have developed various variants, such as generalized multiple-mode OFDM-IM \cite{8425986}, diversity enhancing multiple-mode OFDM-IM\cite{8316922}, and Huffman coded OFDM-IM\cite{8962119} systems to further enhance the system performance from different perspectives. Moreover, a frequency index modulation and a joint code-and-frequency index modulation have been proposed in \cite{7944538} and \cite{8792959}, respectively, tailored for low-complexity and low-power IoT applications. 
In a word, with respective to the conventional modulation schemes, the IM-based schemes can significantly improve the transmission throughput by introducing additional transmission entity to carry information. Nonetheless, how to intelligently incorporate the IM into LoRa modulation to achieve more desirable transmission throughput and error performance has not been explored.

With the aforementioned motivation, we conceive a novel frequency-bin-index (FBI) LoRa system in this paper, which can realize high-data-rate transmissions over wireless channels. We develop two different modulation schemes, referred to as {\em scheme I and scheme II}, to implement the proposed FBI-LoRa system. In scheme I, only the SFBs of the LoRa signal are considered as the transmission entity, which are utilized to carry the source information. To ensure the acceptable computational complexity and hardware-friendly implementation of the receiver, we divide the SFBs within a LoRa signal into several groups before the index modulation. In scheme II, both the the combination of SFB group indices and the combination of SFB indices are utilized to carry the source information, which can be viewed as a generalized version of scheme I. Although both schemes achieve desirable transmission throughput and bit-error-rates (BER), scheme II exploits two-dimensional index and thus benefits from better flexibility and error performance compared with scheme I under the same parameter setting.

To verify the superiority of the proposed FBI-LoRa system, we analyze the theoretical BERs and throughput of the two modulation schemes over additive Gaussian white noise (AWGN) and Rayleigh fading channels, which matches the simulated results very well.
Analysis and simulations have illustrated that the proposed FBI-LoRa system significantly improve system throughput compared to the conventional LoRa system at the price of a slight loss in error performance.

The remainder of this paper is organized as follows. The LoRa modulation is briefly introduced in Section~\ref{sect:Basis of LoRa Modulation}. Section~\ref{sect:frequency-bin-index LoRa System} introduces the proposed FBI-aided modulation scheme. Section~\ref{sect:Performance analysis of the FBI-LoRa} carries out the theoretical performance analysis of the proposed system. Section~\ref{sect:Numerical Results and Discussions} presents various numerical results and discussions. Finally, Section~\ref{sect:Conclusions} concludes the paper.

\section{Fundamentals of LoRa Modulation}\label{sect:Basis of LoRa Modulation}
LoRa modulation is derived from CSS, in which the ${i}^{th}$ transmitted symbol ${{b}_{i}}=o\in \left\{ 0,\ldots,{{2}^{SF}}-1 \right\}$ maps to LoRa symbol ${{s}_{i}}$ by shift chirp modulation. Each LoRa symbol contains ${{2}^{SF}}$ chips, where $SF\in \left\{ 7,\ldots,12 \right\}$ is the spreading factor of LoRa modulation. For a LoRa symbol, the frequency of the baseband signal increases linearly from the starting frequency ${{f}_{st}}=\frac{{{B}_{w}}}{{{2}^{SF}}}$ to ${{B}_{w}}$, then the frequency returns to $0$ but still increases linearly until the end of the symbol duration ${{T}_{sym}}={{2}^{SF}}\cdot {{T}_{chip}}$,  where ${{B}_{w}}$ is the bandwidth of LoRa signal and ${{T}_{chip}}=\frac{1}{{{B}_{w}}}$ is the chip duration. Hence, the discrete time baseband signal of LoRa symbol ${{s}_{i}}$ is expressed as
\begin{align}
   {{s}_{i}}\left( k{{T}_{chip}} \right)&=\sqrt{{{E}_{s}}}{{{\bar{u}}}_{0}}\left( k{{T}_{chip}} \right) \nonumber\\
 & =\sqrt{\frac{{{E}_{s}}}{{{2}^{SF}}}}{{e}^{j2\pi \left[ \frac{{{\left( \left( o+k \right)\bmod {{2}^{SF}} \right)}^{2}}}{{{2}^{SF+1}}} \right]}},
 \label{eq:1func}
\end{align}
where ${E_s}$, $k$, and ${\bar u_0}\left( {k{T_{chip}}} \right)$ are symbol energy, index of the sample at time $k{T_{chip}}$, and the basis function of ${s_i}\left( {k{T_{chip}}} \right)$, respectively. The source information is carried by the SFB index \cite{8067462}, and ${s_i}\left( {k{T_{chip}}} \right)$ can be considered as a cyclic shift of $o{T_{chip}}$ for the basic CSS signal. The basic CSS signal can be called as {\em an upchirp signal}, and its frequency linearly increases from $0$ to ${B_w}$ within a symbol duration. Actually, LoRa symbols with different starting frequencies are mutually orthogonal \cite{9018296,9000860,8067462}, which is a salient feature requiring to be considered in the system design. For the received signal ${r_i}\left( {k{T_{chip}}} \right)$ of a frequency-flat and time invariant wireless channel, the correlator output in the LoRa demodulator is written as
\begin{align}
  {\Psi _\Omega } &= \sum\limits_{k = 0}^{{2^{SF}} - 1} {{r_i}\left( {k{T_{chip}}} \right) \cdot \bar u_\Omega ^ * } \left( {k{T_{chip}}} \right) \hfill \nonumber\\
   &= \sum\limits_{k = 0}^{{2^{SF}} - 1} {\left( {\sqrt {{h_\alpha }} {s_i}\left( {k{T_{chip}}} \right) + n\left( {k{T_{chip}}} \right)} \right)}  \cdot \bar u_\Omega ^ * \left( {k{T_{chip}}} \right) \hfill \nonumber\\
   &= \left\{ \begin{gathered}
  \sqrt {{h_\alpha }{E_s}}  + {\varphi _\Omega }{\text{  }}\ \Omega  = o \hfill \\
  {\varphi _\Omega }{\text{                }}\ \ \ \ \ \ \ \ \ \ \ \ \ \ \Omega  \ne o \hfill \\
\end{gathered}  \right., \hfill
\label{eq:2func}
\end{align}
where $\sqrt {{h_\alpha }} $, $n\left( {k{T_{chip}}} \right)$, ${\varphi _\Omega }$, and $ * $ are the complex envelope amplitude \cite{8746470,8392707}, the complex AWGN, the complex Gaussian noise process, and the complex conjugate operation, respectively.\footnote{In particular, if ${h_\alpha } = 1$, then the channel can be considered as an AWGN channel. }  Accordingly, the transmitted symbol ${b_i}$ can be estimated as
\begin{align}
{\hat b_i} = \arg \mathop {\max }\limits_{\Omega  = 0,\ldots,{2^{SF}} - 1} \left( {\left| {{\Psi _\Omega }} \right|} \right),
\label{eq:3func}
\end{align}
where $\left|  \cdot  \right|$ denotes absolute operation. Furthermore, another equivalent method can also be utilized to perform demodulation. In this method, the received signal is first multiplied by downchirp ${\bar u_{down}}\left( {k{T_{chip}}} \right)$, where ${\bar u_{down}}\left( {k{T_{chip}}} \right)$ is expressed by
\begin{align}
{\bar u_{down}}\left( {k{T_{chip}}} \right) = \sqrt {\frac{1}{{{2^{SF}}}}} {e^{ - j2\pi \frac{{{k^2}}}{{{2^{SF + 1}}}}}}.
\label{eq:4func}
\end{align}
Then, the dechirped signal is processed by ${2^{SF}}$-point discrete Fourier transform (DFT) and yields at
\begin{align}
{\Psi _F} = {\rm{DFT}}\left( {{{\br}_i} \odot {{\bar {{\bu}}}_{down}}} \right),
\label{eq:5func}
\end{align}
where ${\bm{\Psi}_F} = \left[ {{\Psi _{F,0}},\ldots{\Psi _{F,\Omega }},\ldots{\Psi _{F,{2^{SF}} - 1}}} \right]$, ${{\br}_i} = \left[ {{r_i}\left( 0 \right),{r_i}\left( {{T_{chip}}} \right),\ldots,{r_i}\left( {\left( {{2^{SF}} - 1} \right){T_{chip}}} \right)} \right]$, ${\bar {{\bu}}_{down}} = \left[ {{{\bar u}_{down}}\left( 0 \right),\ldots,{{\bar u}_{down}}\left( {{T_{chip}}} \right)} \right]$, and $ \odot $ is the Hadamard product operator\cite{Magnus1988Matrix}. Finally, the transmitted symbol ${b_i}$ is estimated as
\begin{align}
{\hat b_i} = \arg \mathop {\max }\limits_{\Omega  = 1,\ldots,{2^{SF}} - 1} \left( {\left| {{\Psi _{F,\Omega }}} \right|} \right).
\label{eq:6func}
\end{align}
\begin{figure*}[!tbp]
\centering\vspace{-3mm}
\subfigure[\hspace{-0.8cm}]{ \label{fig:subfig:1a}
\includegraphics[width=7.09in,height=1.92in]{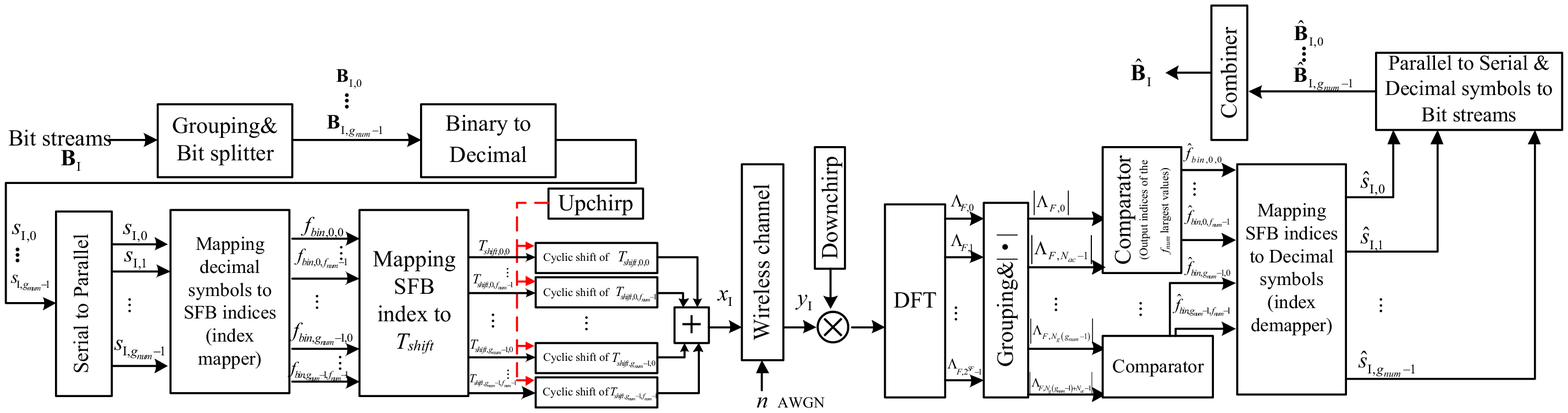}}\vspace{-1.5mm}
\subfigure[\hspace{-0.8cm}]{ \label{fig:subfig:1b}
\includegraphics[width=7.09in,height=2.95in]{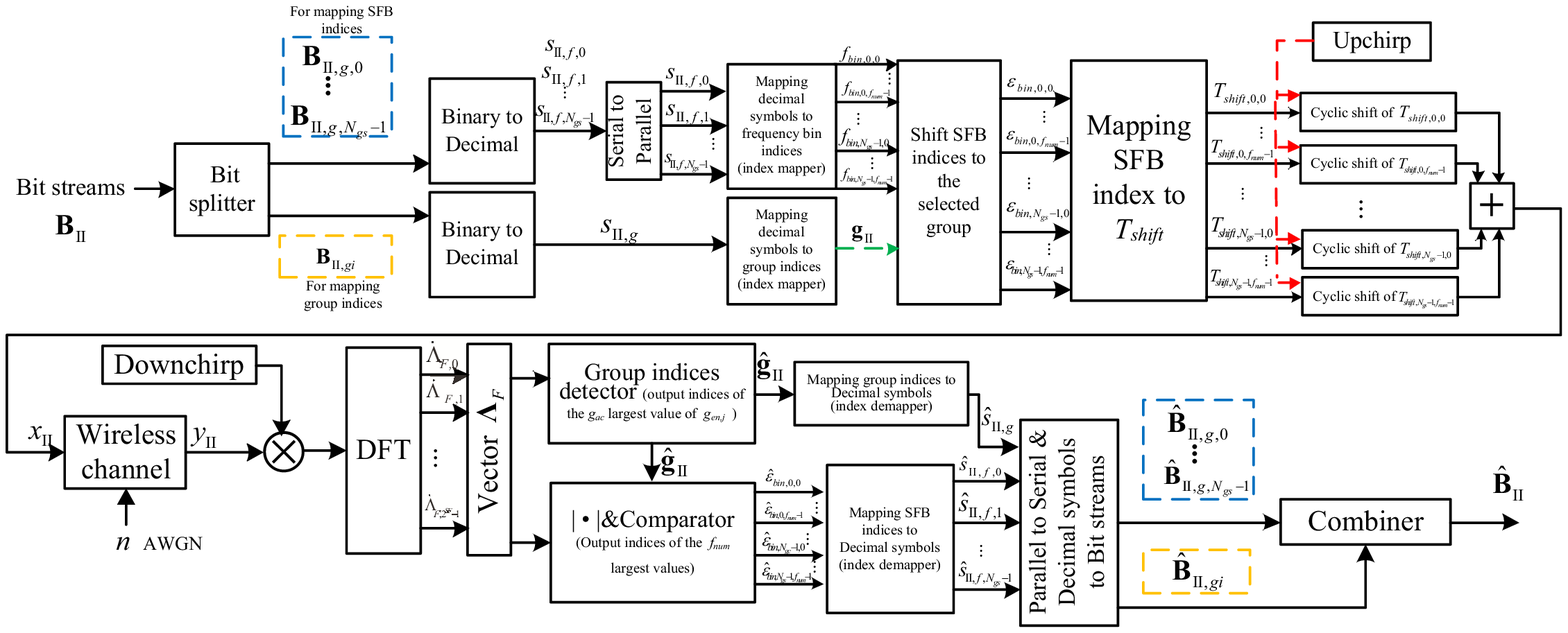}}
\vspace{-0.2cm}
\caption{The block diagram of the transceivers for (a) scheme I and (b) scheme II in the FBI-LoRa system.}
\label{fig:fig1}  
\vspace{-2mm}
\end{figure*}
\section{Frequency-Bin-Index LoRa System}\label{sect:frequency-bin-index LoRa System}
In this section, a new FBI-LoRa system is proposed. In this system, we design two modulation/demodulation schemes (i.e., schemes I and II) for different transmission rate requirements. The block diagram of the transceivers for the two modulation schemes is illustrated in Fig.~\ref{fig:fig1}.
\subsection{Transmitter}\label{sect:Transmitter}
In the proposed FBI-LoRa system, the SFB index is exploited to carry the transmitted information bits. The transmitters of scheme I and scheme II in the FBI-LoRa system are elaborated as follows.

\subsubsection{Scheme I} \label{sect:Scheme I}
The FBI-LoRa signal with spreading factor $SF$ has ${2^{SF}}$ SFB indices $ {0,\ldots,{2^{SF}} - 1} $, which is the same as the conventional LoRa signal. These indices are equally divided into ${g_{num}}$ groups, and thus each group has ${N_g} = \frac{{{2^{SF}}}}{{{g_{num}}}}$ SFB indices. Without loss of generality, we assume that ${N_p}$ SFB indices are utilized to transmit information bits.
In other words, there are ${f_{num}}$ SFB indices in each group in scheme I, where ${f_{num}} = \frac{{{N_p}}}{{{g_{num}}}}$. Therefore, as shown in Fig.~\ref{fig:subfig:1a}, the transmitted signal of scheme I in the proposed FBI-LoRa system is expressed as
\begin{align}
&{x_{\rm{I}}}\left( {k{T_{chip}}} \right) \nonumber\\
&= \sum\limits_{{\tau} = 0}^{{g_{num}} - 1}\!\! {\sum\limits_{{\varsigma} = 0}^{{f_{num}} - 1} \!\!{\sqrt {\frac{{{E_s}}}{{{f_{num}}{g_{num}}}}} } } {e^{j2\pi \left[ {\frac{{{{\left( {\left( {{f_{bin,{\tau},{\varsigma}}} + k} \right)\bmod {2^{SF}}} \right)}^2}}}{{{2^{SF + 1}}}}} \right]}},
\label{eq:7func}
\end{align}
where ${f_{bin,{\tau},{\varsigma}}}$ denotes the $\varsigma^{th}$ (${{\varsigma} = 0,1,\ldots,{f_{num}} - 1}$) selected SFB index in the $\tau^{th}$~(${{\tau} = 0,1,\ldots,{g_{num}} - 1}$) group. Each SFB group can carry ${N_{b,per}} = \left\lfloor {{{\log }_2}\left( {\begin{array}{*{20}{c}}
  {{N_g}} \\
  {{f_{num}}}
\end{array}} \right)} \right\rfloor $ information bits, where $\left( {\begin{array}{*{20}{c}}
  {{A_1}} \\
  {{A_2}}
\end{array}} \right) = \frac{{{A_1}!}}{{{A_2}!\left( {{A_1} - {A_2}} \right)!}}$. Hence, the number of the information bits carried by each symbol in scheme I equals
\begin{align}
{{N}_{tot,\rm{I}}}={{g}_{num}}\left\lfloor {{\log }_{2}}\left( \begin{matrix}
   {{2}^{SF}}/{{g}_{num}}  \\
   {{f}_{num}}  \\
\end{matrix} \right) \right\rfloor ,
 \label{eq:datarate1}
 \end{align}
where $\left\lfloor  \cdot  \right\rfloor $ is the floor function.
For a given $SF$, the data rate of scheme I can be changed by adjusting $g_{num}$ and $f_{num}$.

At the transmitter of scheme I, the overall information-bit stream ${\bB}_{\rm{I}}$ of length ${N_{tot,{\rm{I}}}}$ is divided into ${g_{num}}$ bit streams ${{\bB}_{{\rm{I}},0}},\ldots,{{\bB}_{{\rm{I}},{\tau}}},\ldots,{{\bB}_{{\rm{I}},{g_{num}} - 1}}$ evenly, and then ${{\bB}_{{\rm{I}},{\tau}}}$ utilizes an index mapper to select ${f_{num}}$ SFB indices in the $\tau^{th}$ group.\footnote{In practice, the bit stream ${{\bB}_{{\bf{I}},{\tau}}}$ is first converted to decimal symbol ${s_{{\rm{I,}}{\tau}}}$ utilizing a binary-to-decimal converter, and then an index mapper is utilized for SFB index selection. To be concise, we will omit this step in the description.} Notably, each group is independent of each other, hence the average error probability of each bit stream ${{\bB}_{{\rm{I}},{\tau}}}$ is the same. Within each SFB group, the SFB index is renumbered as $0,\ldots,{N_g} - 1$. In this paper, we defined ${f_{select}} \in \{0,\ldots,{N_g} - 1\}$ as the $\varsigma^{th}$ selected index in the $\tau^{th}$ group, then its corresponding index within the overall LoRa signal equals ${f_{bin,{\tau},{\varsigma}}} = {\tau}{g_{num}} + {f_{select}} \in \{0,\ldots,{2^{SF}} - 1\}$.

\subsubsection{Scheme II} \label{sect:Scheme II}
Different from scheme I, scheme II utilizes not only the SFB index but also the group index to modulate the source information. In this scheme,
we assume that ${N_{gs}}$ ($1 \leqslant {N_{gs}} < {g_{num}}$) groups of SFBs are utilized to carry information bits and that selected group-index vector is ${{\bg}_{{\rm{II}}}} = \left\{ {{g_{{\rm{II}}}}\!\!\left( 0 \right),\ldots,{g_{{\rm{II}}}}\!\!\left( {{\xi}} \right),\ldots,{g_{{\rm{II}}}}\!\!\left( {{N_{gs}} - 1} \right)} \right\}$~$({0 \!\leqslant \!{g_{{\rm{II}}}}\left( {{\xi}} \right) \!\leqslant \!{g_{num}}\!\! -\!\! 1})$. Moreover, within each group of scheme II, ${f_{num}} = \frac{{{N_p}}}{{{N_{gs}}}}$ indices of SFBs are selected for carrying information bits. Thus, as  in Fig.~\ref{fig:subfig:1b}, the transmitted signal of scheme II in the proposed FBI-LoRa system is written as
\begin{align}
&{x_{{\rm{II}}}}\left( {k{T_{chip}}} \right) \nonumber\\
&= \sum\limits_{{\xi} = 0}^{{N_{gs}} - 1} {\sum\limits_{{\gamma} = 0}^{{f_{num}} - 1} {\sqrt {\frac{{{E_s}}}{{{f_{num}}{N_{gs}}}}} {e^{j2\pi \left[ {\frac{{{{\left( {\left( {{{\varepsilon}_{bin,{\xi},{\gamma}}} + k} \right)\bmod {2^{SF}}} \right)}^2}}}{{{2^{SF + 1}}}}} \right]}}} } ,
\label{eq:8func}
\end{align}
where ${\varepsilon_{bin,{\xi},{\gamma}}}$ denotes the $\gamma^{th}$ (${{\gamma} = 0,1,\ldots,{f_{num}} - 1}$) selected SFB index in the group ${g_{{\rm{II}}}}\left( {{\xi}} \right)$. As in scheme I, the SFBs in each group can carry ${N_{b,per}}$ information bits in scheme II, while the combination of group indices can carry additional ${{N}_{b,gi}}$ bits, where ${{N}_{b,gi}}=\left\lfloor {{\log }_{2}}\left( \begin{matrix}
   {{g}_{num}}  \\
   {{N}_{gs}}  \\
\end{matrix} \right) \right\rfloor $.
Thus, the number of the information bit carried by each transmitted symbol in scheme II is
\begin{align}
{{N}_{tot,\rm{II}}}={{N}_{gs}}\left\lfloor {{\log }_{2}}\left( \begin{matrix}
   {{2}^{SF}}\text{/}{{g}_{num}}  \\
   {{f}_{num}}  \\
\end{matrix} \right) \right\rfloor +\left\lfloor {{\log }_{2}}\left( \begin{matrix}
   {{g}_{num}}  \\
   {{N}_{gs}}  \\
\end{matrix} \right) \right\rfloor .
\label{eq:datarate2}
\end{align}
For a given $SF$, the data rate of scheme II can be changed by adjusting $g_{num}$, $f_{num}$, and ${N}_{gs}$.

At the transmitter of scheme II, the overall information-bit stream ${{\bB}_{{\rm{II}}}}$ with length ${N_{tot,{\rm{II}}}}$ is split into a bit stream ${{\bB}_{{\rm{II}},g}}$ with length ${N_{gs}}{N_{b,per}}$ and a bit stream ${\bB_{{\rm{II}},gi}}$ with length ${N_{b,gi}}$, where ${{\bB}_{{\rm{II}},gi}}$ utilizes the index mapper to select ${N_{gs}}$ SFB groups ${g_{{\rm{II}}}}\left( 0 \right),\ldots,{g_{{\rm{II}}}}\left( {{\xi}} \right),\ldots,{g_{{\rm{II}}}}\left( {{N_{gs}} - 1} \right)$. Further, ${{\bB}_{{\rm{II}},g}}$ is divided equally into ${N_{gs}}$ bit streams ${{\bB}_{{\rm{II}},g,0}},\ldots,{{\bB}_{{\rm{II}},g,{\xi}}},\ldots,{{\bB}_{{\rm{II}},g,{N_{gs}} - 1}}$, where ${{\bB}_{{\rm II},g,{\xi}}}$ utilizes the index mapper to select ${f_{num}}$ SFB indices in the group ${g_{{\rm{II}}}}\left( {{\xi}} \right)$ (${{\xi} = 0,1,\ldots,{N_{gs}} - 1}$). It is worth mentioning that the information bits ${{\bB}_{{\rm{II}},gi}}$ carried by the group index have a higher priority than the information bits ${{\bB}_{{\rm{II}},g}}$ because the transmitted power of the former is much larger than the latter. Assuming that ${\varepsilon_{select}}$ is the $\gamma^{th}$ selected index within the group ${g_{{\rm{II}}}}\left( {{\xi}} \right)$, its corresponding index ${\varepsilon_{bin,{\xi},{\gamma}}}$ within the overall LoRa signal becomes ${\varepsilon_{bin,{\xi},{\gamma}}} = {g_{{\rm{II}}}}\left( {{\xi}} \right){g_{num}} + {\varepsilon_{select}} \in ({0,\ldots,{2^{SF}} - 1})$. The details of the SFB index mapping rule will be introduced in Sect.~II-C.

\subsection{Receiver}\label{sect:Receiver}
In the proposed FBI-LoRa system, the received signal of a flat and quasi-static fading channel is given by
\begin{align}
{y_{{\rm{SC}}}}\left( {k{T_{chip}}} \right) = \sqrt \alpha  {x_{\rm{SC}}}\left( {k{T_{chip}}} \right) + n\left( {k{T_{chip}}} \right),
\label{eq:9func}
\end{align}
where $\sqrt \alpha  $ is the magnitude of fading channel coefficient \cite{8746470,8392707}, $n\left( {k{T_{chip}}} \right)$ is the complex AWGN with variance ${N_0}/2$ per dimension, and the subscript ``SC'' in ${y_{{\rm{SC}}}}$ and ${x_{{\rm{SC}}}}$ is utilized to denote either ``scheme I''  or ``scheme II'', i.e., ${\rm{SC}} \in \left\{ {{\rm{I}},{\rm{II}}} \right\}$. Without loss of generality, the chip time ${T_{chip}}$ is equal to 1.

\subsubsection{Scheme I} \label{sect:Scheme I1}

First, a ${2^{SF}}$-point DFT is performed on the received signal after being dechirped, one can obtain \cite{8903531}
\begin{align}
  {\Lambda _{F,v}} &= \frac{1}{{\sqrt {{2^{SF}}} }}\sum\limits_{k = 0}^{{2^{SF}} - 1} {{{\tilde y}_{\rm{I}}}\left( k \right){e^{ - j2\pi \frac{{kv}}{{{2^{SF}}}}}}}  \hfill \nonumber\\
  & = \left\{ \begin{gathered}
  \sqrt {\frac{{\alpha {E_s}}}{{{f_{num}}{g_{num}}}}}  + {\phi _v}{\text{      }}v = {f_{bin,{\tau},{\varsigma}}} \hfill \\
  {\phi _v}\ \ \ \ \ \ \ \ \ \ \ \ \ \ \ \ \ \ \ \ v \ne {f_{bin,{\tau},{\varsigma}}} \hfill \\
\end{gathered}  \right., \hfill
\label{eq:10func}
\end{align}
where ${\tilde y_{\rm{I}}}\left( k \right)\!\! =\!\! {y_{\rm{I}}}\left( k \right){\bar u_{down}}\left( k \right)$. In particular,
the dechirped signal vector can be written as ${\bm{\Lambda} _F} \!\!=\!\! \left\{ {\bm{\Lambda} _F^0,\ldots,\bm{\Lambda} _F^{{\tau}},\ldots,\bm{\Lambda} _F^{{g_{num}} - 1}} \right\}\!\! =\!\! \left\{ {{\Lambda _{F,0}},\ldots,{\Lambda _{F,v}},\ldots,{\Lambda _{F,{2^{SF}} - 1}}} \right\}$, where $\bm{\Lambda} _F^{{\tau}}$ is the dechirped signal corresponding to the $\tau^{th}$ SFB group transmitted over the channel.
Actually, in $\bm{\Lambda} _F^{{\tau}}$, only ${N_{ac}}$ dechirped samples (${v = {\tau}{N_g},{\tau}{N_g} + 1,\ldots,{\tau}{N_g} + {N_{ac}} - 1}$) need to be exploited for demodulation. Therefore, the selected SFB indices can be estimated as
\begin{align}
  {{\hat {\bf{f}}}_{bin,{\tau}}} &= \left\{ {{{\hat f}_{bin,{\tau},0}},\ldots,{{\hat f}_{bin,{\tau},{f_{num}} - 1}}} \right\} \hfill \nonumber\\
  & = \arg \mathop {{{\max }^{{f_{num}}}}}\limits_{v = {\tau}{N_g},\ldots,{\tau}{N_g} + {N_{ac}} - 1} \left( {\left| {{\Lambda _{F,v}}} \right|} \right), \hfill
  \label{eq:11func}
\end{align}
where $\arg \mathop {{{\max }^{{n_1}}}}\limits_{V \in {{\bV}_1}} \left( {{A_V}} \right)$ denotes the output index $V$ corresponding to the ${n_1}$ largest values in the set ${\bA} = \left\{ {{A_V}|V \in {{\bV}_1}} \right\}$, and ${N_{ac}}$ is the minimum value that satisfies the following inequation
\begin{align}
{2^{{N_{b,per}}}} \le \left( {\begin{array}{*{20}{c}}
{{N_{ac}}}\\
{{f_{num}}}
\end{array}} \right) \le \left( {\begin{array}{*{20}{c}}
{{N_g}}\\
{{f_{num}}}
\end{array}} \right).
\label{eq:inequation1}
\end{align}
Finally, ${\hat {\bf{f}}_{bin,{\tau}}}$ utilizes the index demapper to get ${\hat {{\bB}}_{{\rm{I}},{\tau}}}$, and one can employ the combiner to get the estimated information bits ${\hat{ {\bB}}_{\rm{I}}}$.
\subsubsection{Scheme II} \label{sect:Scheme II1}
As in scheme I, the ${2^{SF}}$-point DFT is first performed on the received signal, i.e.,
\begin{align}
  {{\dot \Lambda }_{F,v}} &= \frac{1}{{\sqrt {{2^{SF}}} }}\sum\limits_{k = 0}^{{2^{SF}} - 1} {{{\tilde y}_{{\rm{II}}}}\left( k \right){e^{ - j2\pi \frac{{kv}}{{{2^{SF}}}}}}}  \hfill \nonumber\\
  & = \left\{ \begin{gathered}
  \sqrt {\frac{{\alpha {E_s}}}{{{f_{num}}{N_{gs}}}}}  + {\phi _v}{\text{      }}v = {{\varepsilon}_{bin,{\xi},{\gamma}}} \hfill \\
  {\phi _v}\ \ \ \ \ \ \ \ \ \ \ \ \ \ \ \ \ \ \ v \ne {{\varepsilon}_{bin,{\xi},{\gamma}}} \hfill \\
\end{gathered}  \right., \hfill
\label{eq:12func}
\end{align}
where ${\tilde y_{{\rm{II}}}}\left( k \right) = {y_{{\rm{II}}}}\left( k \right){\bar u_{down}}\left( k \right)$. Then, we employ a frequency-domain-energy-based detection to estimate the indices of the selected SFB groups. Let dechirped
signal vector ${\dot {\bm{\Lambda}} _F} = \left\{ {\dot{ \bm{\Lambda}} _F^0,\ldots,\dot {\bm{\Lambda}} _F^\varphi ,\ldots,\dot{\bm{ \Lambda}} _F^{{g_{num}} - 1}} \right\}$, where $\dot {\bm{\Lambda}} _F^\varphi  = \left\{ {{{\dot \Lambda }_{F,\varphi {N_g}}},{{\dot \Lambda }_{F,\varphi {N_g} + 1}},\ldots,{{\dot \Lambda }_{F,{N_g}\left( {\varphi  + 1} \right) - 1}}} \right\}$. In fact, only ${g_{ac}}$ dechirped
singles need to be detected. Hence, the selected SFB group indices can be estimated as
\begin{align}
  {{\hat g}_{{\rm{II}}}} &= \left\{ {{{\hat g}_{{\rm{II}}}}\left( 0 \right),\ldots,{{\hat g}_{{\rm{II}}}}\left( {{N_{gs}} - 1} \right)} \right\} \hfill \nonumber\\
  {\text{    }} &= \arg \mathop {{{\max }^{{N_{gs}}}}}\limits_{\varphi  = 0,1,\ldots,{g_{ac}} - 1} \left( {{g_{en,\varphi }}} \right), \hfill
  \label{eq:13func}
\end{align}
where
\begin{align}
\begin{gathered}
  {g_{en,\varphi }} = \sum\limits_{\chi  = 0}^{{N_{ac}} - 1} {{{\left| {{{\dot \Lambda }_{F,\varphi {N_g} + \chi }}} \right|}^2}}  \hfill \\
   = \left\{\!\!\!\! \begin{gathered}
  \left( \!\!\!\!\begin{gathered}
  \sum\limits_{\begin{subarray}{c}
  {\chi {\text{ = 0}}} \\
  {\varphi {N_g} + \chi  = {f_{bin,{\xi},{\gamma}}}}
\end{subarray}}^{{N_{ac}} - 1}\!\! {{{\left| {\sqrt {\frac{{\alpha {E_s}}}{{{f_{num}}{N_{gs}}}}}  + {\phi _{\varphi {N_g} + \chi }}} \right|}^2}}  \hfill \\
   + \sum\limits_{\begin{subarray}{c}
  {\chi {\text{ = 0}}} \\
  {\varphi {N_g} + \chi  \ne {f_{bin,{\xi},{\gamma}}}}
\end{subarray}}^{{N_{ac}} - 1} {{{\left| {{\phi _{\varphi {N_g} + \chi }}} \right|}^2}{\text{   }}}  \hfill \\
\end{gathered}  \right){\text{                         }}\varphi  = {g_{{\text{II}}}}\left( {{\xi}} \right) \hfill \\
  \sum\limits_{\chi  = 0}^{{N_{ac}} - 1} {{{\left| {{\phi _{\varphi {N_g} + \chi }}} \right|}^2}{\text{                                                           }}\ \ \ \ \ \ \ \ \ \ \ \ \ \ \ \ \ \ \ \ \ \ \ \ \ \ \ \ \ \ \ \ \varphi  \ne {g_{{\text{II}}}}\left( {{\xi}} \right)}  \hfill \\
\end{gathered}  \right., \hfill \\
\end{gathered}
\label{eq:14func}
\end{align}
and ${g_{ac}}$ is the minimum value that satisfies the following inequation
\begin{align}
{2^{{N_{b,gi}}}} \leqslant \left( {\begin{array}{*{20}{c}}
  {{g_{ac}}} \\
  {{N_{gs}}}
\end{array}} \right) \leqslant \left( {\begin{array}{*{20}{c}}
  {{g_{num}}} \\
  {{N_{gs}}}
\end{array}} \right).
\label{eq:inequation2}
\end{align}
Moreover, we utilize the index demapper to obtain ${\hat {{\bB}}_{{\rm{II}},gi}}$. One can subsequently estimate the selected SFB indices as
\begin{align}
  {{\hat { {{\bE}}}}_{bin,{\xi}}} &= \left\{ {{{\hat { {\varepsilon }}}_{bin,{\xi},0}},\ldots,{{\hat { {\varepsilon}}}_{bin,{\xi},{f_{num}}}}} \right\} \hfill \nonumber\\
  &= \arg \mathop {{{\max }^{{f_{num}}}}}\limits_{v = {\xi}{N_g},\ldots,{\xi}{N_g} + {N_{ac}} - 1} \left( {\left| {{{\dot \Lambda }_{F,v}}} \right|} \right). \hfill
  \label{eq:15func}
\end{align}
Finally, ${\hat { {{\bE}}}_{bin,{\xi}}}$ utilizes the index demapper to yield ${\hat {{\bB}}_{{\rm{II}},g,{\xi}}}$. Hence, one can employ the combiner to obtain the estimated information bits ${\hat {{\bB}}_{{\rm{II}}}}$.

\begin{figure*}[tbp]
\centering\vspace{-3mm}
\subfigure[\hspace{-0.8cm}]{ \label{fig:subfig:2a}
\includegraphics[width=7in,height=1.0831in]{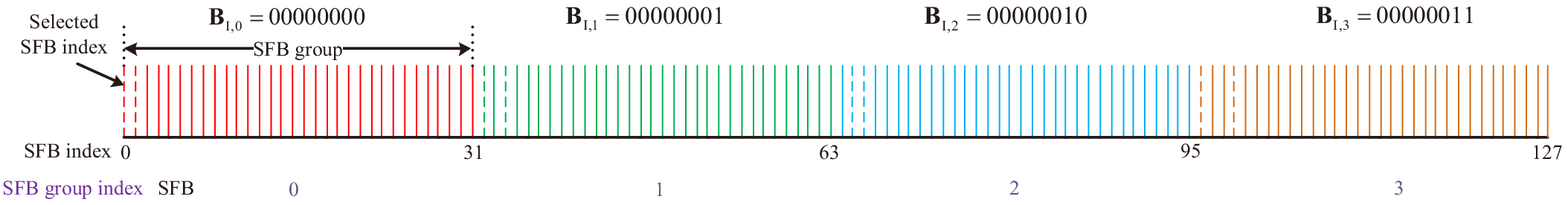}}\vspace{-1.5mm}
\subfigure[\hspace{-0.8cm}]{ \label{fig:subfig:2b}
\includegraphics[width=7in,height=1.135in]{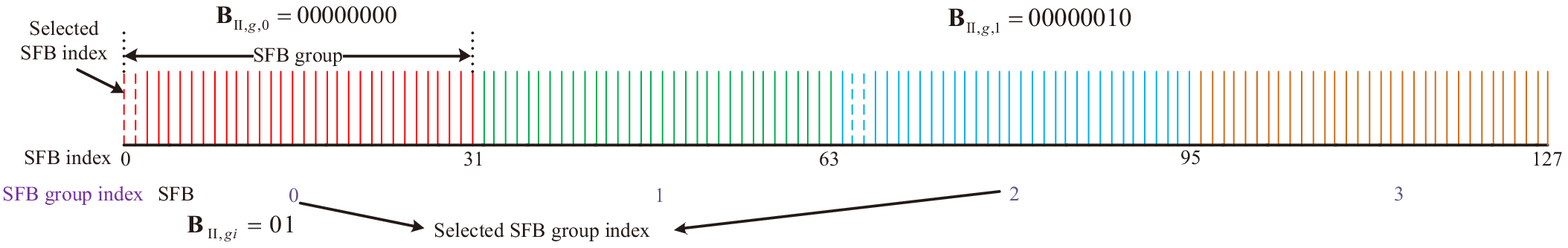}}
\vspace{-0.2cm}
\caption{Examples for the SFB index structures of (a) scheme I and (b) scheme II in the FBI-LoRa system.}
\label{fig:fig2}  
\vspace{-2mm}
\end{figure*}
\subsection{SFB Index Mapper}\label{sect:SFB Index mapper}
It is worth mentioning that the proposed FBI-LoRa system can be implemented without the division of SFB indices (i.e. set  ${g_{num}} = 1$). Unfortunately, the number of the combinations of SFBs in this case, i.e., the value of $\left( {\begin{array}{*{20}{c}}
  {{{\text{2}}^{SF}}} \\
  {{f_{num}}}
\end{array}} \right)$, may be very large, which makes the implementation of the proposed FBI-LoRa system extremely difficult. The division of the SFB indices of a LoRa signal is beneficial to not only the friendly implementation of the FBI-LoRa system, but also flexible realization of diverse data-rate requirements (e.g., scheme I and scheme II). In the existing IM-based communication systems, there are two common index mapping methods: the look-up table method and the combinational method \cite{6587554}. Although table lookup is a simple method, designing and maintaining such a table can be very difficult when the combination number $\left( {\begin{array}{*{20}{c}}
  {{N_g}} \\
  {{f_{num}}}
\end{array}} \right)$ takes a relatively large value, thus making the system difficult to be implemented. Thereby, the combinatorial method is considered in the proposed FBI-LoRa system for mapping the SFB indices.

Here, we briefly introduce the combinatorial method. For any natural number $Z \in \left[ {0,\left( {\begin{array}{*{20}{c}}
  {{m_1}} \\
  {{m_2}}
\end{array}} \right) - 1} \right]$, the combinatorial method can map $Z$ into a strictly monotonically decreasing sequence with length ${m_2}$. In other words, for fixed values of ${m_1}$ and ${m_2}$, $Z$ can be mapped to a sequence $R = \left\{ {{d_{{m_2}}},\ldots,{d_1}} \right\}$ with length ${m_2}$, where ${d_{{m_1}}} >  \cdots  > {d_1} \geqslant 0$ and ${d_1},\ldots,{d_{{m_2}}} \in \left\{ {1,\ldots,{m_1} - 1} \right\}$. The elements of the sequence $R$ can be obtained via the following equation
\begin{align}
Z=\left( \begin{matrix}
   {{d}_{{{m}_{2}}}}  \\
   {{m}_{2}}  \\
\end{matrix} \right)+\cdots +\left( \begin{matrix}
   {{d}_{2}}  \\
   2  \\
\end{matrix} \right)+\left( \begin{matrix}
   {{d}_{1}}  \\
   1  \\
\end{matrix} \right).
\label{eq:16func}
\end{align}
As an example, for ${{m}_{1}}=8$, ${{m}_{2}}=3$, and $\left( \begin{matrix}
   8  \\
   3  \\
\end{matrix} \right)=56$, the corresponding sequences $R$ can be obtained as
\[\begin{matrix}
  55=\left( \begin{matrix}
   7  \\
   3  \\
\end{matrix} \right)+\left( \begin{matrix}
   6  \\
   2  \\
\end{matrix} \right)+\left( \begin{matrix}
   5  \\
   1  \\
\end{matrix} \right)\to R=\left\{ 7,6,5 \right\}, \\
  54=\left( \begin{matrix}
   7  \\
   3  \\
\end{matrix} \right)+\left( \begin{matrix}
   6  \\
   2  \\
\end{matrix} \right)+\left( \begin{matrix}
   4  \\
   1  \\
\end{matrix} \right)\to R=\left\{ 7,6,4 \right\}, \\
  \vdots  \\
  23=\left( \begin{matrix}
   6  \\
   3  \\
\end{matrix} \right)+\left( \begin{matrix}
   3  \\
   2  \\
\end{matrix} \right)+\left( \begin{matrix}
   0  \\
   1  \\
\end{matrix} \right)\to R=\left\{ 6,3,0 \right\}, \\
  22=\left( \begin{matrix}
   6  \\
   3  \\
\end{matrix} \right)+\left( \begin{matrix}
   2  \\
   2  \\
\end{matrix} \right)+\left( \begin{matrix}
   1  \\
   1  \\
\end{matrix} \right)\to R=\left\{ 6,2,1 \right\}, \\
  \vdots  \\
  1=\left( \begin{matrix}
   3  \\
   3  \\
\end{matrix} \right)+\left( \begin{matrix}
   1  \\
   2  \\
\end{matrix} \right)+\left( \begin{matrix}
   0  \\
   1  \\
\end{matrix} \right)\to R=\left\{ 3,1,0 \right\}, \\
  0=\left( \begin{matrix}
   2  \\
   3  \\
\end{matrix} \right)+\left( \begin{matrix}
   1  \\
   2  \\
\end{matrix} \right)+\left( \begin{matrix}
   0  \\
   1  \\
\end{matrix} \right)\to R=\left\{ 2,1,0 \right\}. \\
\end{matrix}\]
For all ${{m}_{1}}$ and ${{m}_{2}}$, the generation method of the sequence $R$ can be simply described as follows: i) select the largest ${{d}_{{{m}_{2}}}}$ to satisfy $\left( \begin{matrix}
   {{d}_{{{m}_{2}}}}  \\
   {{m}_{2}}  \\
\end{matrix} \right)\le Z$, ii) select the largest ${{d}_{{{m}_{2}}-1}}$ to satisfy $\left( \begin{matrix}
   {{d}_{{{m}_{2}}-1}}  \\
   {{m}_{2}}-1  \\
\end{matrix} \right)$, and so on, until the sequence $R$ with length ${{m}_{2}}$ is yielded. In the proposed FBI-LoRa system, the information bits need to be converted into decimal symbols at the transmitter, then the SFB indices are selected by the combinatorial method.
At the receiver, the combination of the selected indices is estimated, and then the decimal symbol is obtained through Eq.~\eqref{eq:16func}.
Fig.~\ref{fig:fig2} gives two examples to illustrate the SFB index structures for scheme I and scheme II.

\section{Performance Analysis}\label{sect:Performance analysis of the FBI-LoRa}
\subsection{BER Performance}\label{sect:System Error Probability Analysis}

\subsubsection{Scheme I} \label{sect:Scheme I}
For scheme I, because the detection of the selected SFB indices in each group is obviously independent of one another, the error probabilities of the information bits ${{{\bB}}_{\rm{I},{\tau}}}$ carried in each group are identical. Similar to the conventional IM-based communication systems, the BER of scheme I is expressed as \cite{IM2017,8425986,6587554}
\begin{align}
  {P_{b,{\rm{I}}}} =& \sum\limits_{{{\bf{f}}_{bin,{B_{{\rm{I}},{\tau}}}}}} {\sum\limits_{{{\hat {\bf{f}}}_{bin,{B_{{\rm{I}},{\tau}}}}}} {\left[ \begin{gathered}
  \Pr \left( {{{\bf{f}}_{bin,{{\bB}_{{\rm{I}},{\tau}}}}} \to {{\hat {\bf{f}}}_{bin,{{\bB}_{{\rm{I}},{\tau}}}}}} \right) \hfill \nonumber\\
   \times {n_{er}}\left( {{{\bf{f}}_{bin,{{\bB}_{{\rm{I}},{\tau}}}}},{{\hat {\bf{f}}}_{bin,{\bf{B}_{{\rm{I}},{\tau}}}}}} \right) \hfill \nonumber\\
\end{gathered}  \right]} }  \hfill \nonumber\\
  & \times \frac{1}{{n_{\bf{f},{\rm{I}}}}{N_{b,per}}}, \hfill
   \label{eq:17func}
\end{align}
where ${n_{\bf{f},{\rm{I}}}} = {2^{{N_{b,per}}}}$ is the number of the possible realizations of ${{\bf{f}}_{bin,{{\bB}_{{\rm{I}},{\tau}}}}}$, ${{\bf{f}}_{bin,{{\bB}_{{\rm{I}},{\tau}}}}}$ denotes the set of SFB indices selected by the information bit stream ${{\bB}_{{\rm{I}},{\tau}}}$, ${\bF} \to \hat {{\bF}}$ denotes that ${\bF}$ is erroneously detected as $\hat {{\bF}}$, $\Pr \left(  \cdot  \right)$ denotes the probability of an event, and ${n_{er}}\left( {{\bF},\hat {{\bF}}} \right)$ indicates the number of error bits when ${\bF}$ is erroneously detected as $\hat {{\bF}}$.
The conditional pairwise error probability $\Pr \left( {{{\bf{f}}_{bin,{{\bB}_{{\rm{I}},{\tau}}}}} \to {{\hat {{\bf{f}}}}_{bin,{{\bB}_{{\rm{I}},{\tau}}}}}|\alpha } \right)$ depends on ${k_{i,er}}$, which is the erroneously detected number of SFB indices. Specifically, $\Pr \left( {{{\bf{f}}_{bin,{{\bB}_{{\rm{I}},{\tau}}}}} \to {{\hat {{\bf{f}}}}_{bin,{{\bB}_{{\rm{I}},{\tau}}}}}|\alpha } \right)$ can be expressed as
\begin{align}
&\Pr \left( {{{\bf{f}}_{bin,{{\bB}_{{\rm{I}},{\tau}}}}}  \to  {{\hat {{\bf{f}}}}_{bin,{{\bB}_{{\rm{I}},{\tau}}}}}|\alpha } \right)\nonumber\\
&= \left({P_{ie,{\rm{I}}|\alpha }}\right)^{{k_{i,er}}}{\left( {1 - {P_{ie,{\rm{I}}|\alpha }}} \right)^{{f_{num}} - {k_{i,er}}}},
   \label{eq:18func}
\end{align}
where ${P_{ie,{\rm{I}}|\alpha }}$ represents the conditional error probability of the SFB index detection.
In this paper, because Rayleigh fading channel is considered, $\sqrt \alpha $ is a Rayleigh random variable and thus $\alpha $ follows a Chi-square distribution with one degree of freedom and an expected normalized channel power of one \cite{8746470,8392707}. The  probability density function (PDF) of $\alpha $ can be expressed as ${f_\alpha }\left( \alpha  \right) = {e^{ - \alpha }}$.
Hence, one can obtain
\begin{align}
  \Pr \left( {{{\bf{f}}_{bin,{{\bB}_{{\rm{I}},{\tau}}}}} \to {{\hat {{\bf{f}}}}_{bin,{{\bB}_{{\rm{I}},{\tau}}}}}} \right)\!\!=&\!\! \int_0^\infty  {\Pr \left( {{{\bf{f}}_{bin,{{\bB}_{{\rm{I}},{\tau}}}}} \to {{\hat {{\bf{f}}}}_{bin,{{\bB}_{{\rm{I}},{\tau}}}}}|\alpha } \right)}  \hfill \nonumber\\
   &\times {{f}_\alpha }\left( \alpha  \right)d\alpha . \hfill
\label{eq:pepresult}
\end{align}
 Furthermore, the symbol error rate (SER) of scheme I can be written as
\begin{align}
{P_{s,{\rm{I}}}} = \int_0^\infty  {\left[ {1 - {{\left( {1 - {P_{ie,{\rm{I}}|\alpha}}} \right)}^{{f_{num}}{g_{num}}}}} \right]}  \times {f_\alpha }\left( \alpha  \right)d\alpha .
\label{eq:19func}
\end{align}

\subsubsection{Scheme II} \label{sect:Scheme II}
In scheme II, the ${N_{b,gi}}$ information bits ${{\bB}_{{\rm{II}},gi}}$ are carried by group indices, while the ${N_{gs}}{N_{b,per}}$ information bits ${{\bB}_{{\rm{II}},g}}$ are carried by SFB indices. The BER of total information bits for scheme II is expressed as
\begin{align}
{P_{b,{\rm{II}}}} = \frac{{{P_{b,{\rm{II}},gi}}{N_{b,gi}} + {P_{b,{\rm{II}},g}}{N_{gs}}{N_{b,per}}}}{{{N_{b,gi}} + {N_{gs}}{N_{b,per}}}},
\label{eq:20func}
\end{align}
where ${P_{b,{\rm{II}},gi}}$ and ${P_{b,{\rm{II}},g}}$ are the BERs of ${{\bB}_{{\rm{II}},gi}}$ and ${{\bB}_{{\rm{II}},g}}$, respectively. Analogous to Eq.~(\ref{eq:17func}), ${P_{b,{\rm{II}},gi}}$ is given by
\begin{align}
  {P_{b,{\rm{II,}}gi}} =& \sum\limits_{{{\bf{f}}_{bin,{{\bg}_{{\rm{II}}}}}}} {\sum\limits_{{{\hat {{\bf{f}}}}_{bin,{{\bg}_{{\rm{II}}}}}}} {\left[ \begin{gathered}
  \Pr \left( {{{\bf{f}}_{bin,{{\bg}_{{\rm{II}}}}}} \to {{\hat {{\bf{f}}}}_{bin,{{\bg}_{{\rm{II}}}}}}} \right) \hfill \nonumber\\
   \times {n_{er}}\left( {{{\bf{f}}_{bin,{{\bg}_{{\rm{II}}}}}},{{\hat {{\bf{f}}}}_{bin,{{\bg}_{{\rm{II}}}}}}} \right) \hfill \nonumber\\
\end{gathered}  \right]} }  \hfill \nonumber\\
   &\times\frac{1}{{n_{{\bf{f}},gi,{\rm{II}}}}{N_{b,gi}}} \hfill
   \label{eq:21func}
\end{align}
where ${n_{{\bf{f}},gi,{\rm{II}}}} = {2^{{N_{b,gi}}}}$ is the number of the possible realizations of $ {{\bf{f}}_{bin,{{\bg}_{{\rm{II}}}}}}$, $ {{\bf{f}}_{bin,{{\bg}_{{\rm{II}}}}}}$ denotes the set of SFB group indices selected by the information bit stream ${{\bB}_{{\rm{II}},gi}}$.
The conditional pairwise error probability $\Pr \left( {{{\bf{f}}_{bin,{{\bg}_{{\rm{II}}}}}} \to {{\hat {{\bf{f}}}}_{bin,{{\bg}_{{\rm{II}}}}}}|\alpha } \right)$ depends on ${k_{gi,er}}$, which is the erroneously detected number of SFB group indices. Specifically, $\Pr \left( {{{\bf{f}}_{bin,{{\bg}_{{\rm{II}}}}}} \to {{\hat {{\bf{f}}}}_{bin,{{\bg}_{{\rm{II}}}}}}|\alpha } \right)$ can be expressed as
\begin{align}
&\Pr \left( {{{\bf{f}}_{bin,{{\bg}_{{\rm{II}}}}}}  \to  {{\hat {{\bf{f}}}}_{bin,{{\bg}_{{\rm{II}}}}}}{{|}}\alpha } \right) \nonumber\\
& =   \left({P_{gie,{\rm{II}} |\alpha}}\right)^{{k_{gi,er}}}{\left( {1  -  {P_{gie,{\rm{II}|\alpha} }}} \right)^{{N_{gs}}  -  {k_{gi,er}}}},
\label{eq:22func}
\end{align}
where ${P_{gie,{\rm{II}|\alpha}}}$ denotes the conditional error probability of the SFB group index detection. Afterwards, one can get
\begin{align}
\begin{gathered}
  \Pr \left( {{{\bf{f}}_{bin,{{\bg}_{{\rm{II}}}}}} \to {{\hat {{\bf{f}}}}_{bin,{{\bg}_{{\rm{II}}}}}}} \right) = \int_0^\infty  {\Pr \left( {{{\bf{f}}_{bin,{{\bg}_{{\rm{II}}}}}} \to {{\hat {{\bf{f}}}}_{bin,{{\bg}_{{\rm{II}}}}}}{{|}}\alpha } \right)}  \hfill \\
   \ \ \ \ \ \ \ \ \ \ \ \ \ \ \ \ \ \ \ \ \ \ \ \ \ \ \ \ \ \ \,\,\,\times {{f}_\alpha }\left( \alpha  \right)d\alpha , \hfill \\
\end{gathered}
\label{eq:pepresult3}
\end{align}

Likewise, ${P_{b,{\rm{II}},g}}$ can be expressed as
\begin{align}
  {{P}_{b,\text{II,}g}}&=\sum\limits_{{{\bf{f}}_{bin,{{g}_{\text{II}}}}}}{\sum\limits_{{{{\hat{\bf{f}}}}_{bin,{{g}_{\rm{II}}}}}}{\left[ \begin{gathered}
   \Pr \left( {{\bf{f}}_{bin,{{\bB}_{\rm{II},g,\xi }}}}\to {{{\hat {\bf{f}}}}_{bin,{{B}_{\text{II},g,\xi }}}} \right) \nonumber\\
  \times {{n}_{er}}\left( {{\bf{f}}_{bin,{{\bB}_{\rm{II},g,\xi }}}},{{{\hat{\bf{f}}}}_{bin,{{\bB}_{\rm{II},g,\xi }}}} \right) \nonumber\\
\end{gathered} \right]}} \nonumber\\
 &\ \ \ \ \times \frac{({{N}_{gs}}-{{k}_{gi,er}})}{{{n}_{{\bf{f}},i,\rm{II}}}{{N}_{b,per}}{{N}_{gs}}}+\frac{1}{2}\times \frac{{{k}_{gi,er}}}{{{N}_{gs}}},
 \label{eq:23func}
\end{align}
where ${n_{{\bf{f}},i,{\rm{II}}}} = {2^{{N_{b,per}}}}$ is the number of the possible realizations of ${{\bf{f}}_{bin,{{\bB}_{{\rm{II}},g,{\xi}}}}}$, ${{\bf{f}}_{bin,{{\bB}_{{\rm{II}},g,{\xi}}}}}$ denotes the set of SFB group indices selected by the information bit stream ${{\bB}_{{\rm{II}},g,{\xi}}}$. The conditional pairwise error probability $\Pr \left( {{{\bf{f}}_{bin,{{\bB}_{{\rm{II}},g,{\xi}}}}} \to {{\hat {{\bf{f}}}}_{bin,{{\bB}_{{\rm{II}},g,{\xi}}}}}}|\alpha \right)$ depends on ${k_{i,er}}$, given by
\begin{align}
&\Pr\!\! \left( \!{{{\bf{f}}_{bin,{{\bB}_{{\rm{II}},g,{\xi}}}}}\!\! \!\to \!{{\hat {{\bf{f}}}}_{bin,{{\bB}_{{\rm{II}},g,{\xi}}}}}}|\alpha\! \right)\nonumber\\
& =\!\! \left({P_{ie,{\rm{II}|\alpha}}}\right)^{{k_{i,er}}}{\left(\! {1 \!\!-\!\! {P_{ie,{\rm{II}|\alpha}}}}\! \right)^{{f_{num}} \!-\! {k_{i,er}}}},
  \label{eq:24func}
\end{align}
where ${P_{ie,{\rm{II}|\alpha}}}$ represents the conditional detection error probability of the SFB index.
Subsequently, one has
\begin{align}
  &\Pr \left( {{{\bf{f}}_{bin,{{\bB}_{{\rm{II}},g,{\xi}}}}}\!\! \to\!\! {{\hat {{\bf{f}}}}_{bin,{{\bB}_{{\rm{II}},g,{\xi}}}}}} \right) \nonumber\\
  &= \!\!\int_{\text{0}}^\infty  \!\!{\Pr \left(\!\! {{{\bf{f}}_{bin,{{\bB}_{{\rm{II}},g,{\xi}}}}}\!\! \to \!\!{{\hat {{\bf{f}}}}_{bin,{{\bB}_{{\rm{II}},g,{\xi}}}}}|\alpha } \!\!\right)}   \times {{f}_\alpha }\left( \alpha  \right)d\alpha ,  \hfill
\label{eq:pepresult2}
\end{align}
Finally, the SER of scheme II is written as
\begin{align}
&{P_{s,{\rm{II}}}}\nonumber\\
&= \int_0^\infty \!\! {\left[ \!{1 \!-\! {{\left( {1\! -\! {P_{gie,{\rm{II}|\alpha}}}} \right)}^{{N_{gs}}}}{{\left( {1\! -\! {P_{ie,{\rm{II}|\alpha}}}} \right)}^{{f_{num}}{N_{gs}}}}} \!\right]\!\! \times \!\! {f_\alpha }\!\!\left( \!\alpha\!  \right)} d\alpha .
\label{eq:25func}
\end{align}
\subsection{Derivation of ${P_{ie,{\rm{I}|\alpha}}}$, ${P_{ie,{\rm{II}|\alpha}}}$, and ${P_{gie,{\rm{II}|\alpha}}}$}\label{sect:Derivation of p}
\subsubsection{Derivation of ${P_{ie,{\rm{I}|\alpha}}}$ and ${P_{ie,{\rm{II}|\alpha}}}$}
According to Eq.~\eqref{eq:11func}, ${P_{ie,{\rm{I}}|\alpha}}$ can be expressed as
\begin{align}
  {P_{ie,{\rm{I}}|\alpha}} &= \Pr \left[ {\max \left( {\left| {{\Lambda _{F,v}}} \right|} \right) > \left| {{\Lambda _{F,{f_{bin,{\tau},{\varsigma}}}}}} \right|} \right] \hfill \nonumber\\
  & = \Pr \left[ {\mathop {\max }\limits_{v,v \ne {f_{bin,{\tau},{\varsigma}}}} \left( {\left| {{\phi _v}} \right|} \right) > {\beta _\Lambda }} \right], \hfill
  \label{eq:26func}
\end{align}
where ${\beta _\Lambda } = \left| {\sqrt {\frac{{\alpha {E_s}}}{{{f_{num}}{g_{num}}}}}  + {\phi _{{f_{bin,{\tau},{\varsigma}}}}}} \right|$. Conditioned on $\alpha $, ${\beta _\Lambda }$ follows a Rice distribution with the shape parameter of ${K_\Lambda } = \frac{{\alpha {E_s}}}{{{f_{num}}{g_{num}}{N_0}}}$. Let ${\rho _n} = {\max _{v,v \ne {f_{bin,{\tau},{\varsigma}}}}}\left( {\left| {{\phi _v}} \right|} \right)$ depict the maximum value of the ${N_g} - {f_{num}}$ independent and identically distributed Rayleigh random variables $\left| {{\phi _v}} \right|$. Consequently, the cumulative distribution function (CDF) of ${\rho _n}$ is given by
\begin{align}
{F_{{\rho _n}}}\left( {{\rho _n}} \right) = {\left[ {1 - \exp \left( {\frac{{\rho _n^2}}{{{N_0}}}} \right)} \right]^{{N_{ac}} - {f_{num}}}}.
\label{eq:27func}
\end{align}

According to the distribution of ${\beta _\Lambda }$ and Eq.~(\ref{eq:27func}), Eq.~(\ref{eq:26func}) can be further formulated as
\begin{align}
{P_{ie,{\rm{I}}|\alpha }}\!\! =\!\! \int_0^\infty \!\! {\left[ {1 - {{\left[ {1\!\! - \!\!\exp \!\left( \!{ - \frac{{\beta _\Lambda ^2}}{{{N_0}}}} \right)} \right]}^{{N_{ac}} - {f_{num}}}}} \right]} \!\! \times \!\! {f_{{\beta _\Lambda }}}\left( {{\beta _\Lambda }} \right)d{\beta _\Lambda },
  \label{eq:28func}
\end{align}
where ${f_{{\beta _\Lambda }}}\left( {{\beta _\Lambda }} \right)$ is the PDF of ${\beta _\Lambda }$, i.e.,
\begin{align}
{f_{{\beta _\Lambda }}}\left( {{\beta _\Lambda }} \right) = \frac{{2{\beta _\Lambda }}}{{{N_0}}}{I_0}\left( {\frac{{2{S_\Lambda }{\beta _\Lambda }}}{{{N_0}}}} \right){e^{ - \frac{{\beta _\Lambda ^2 + S_\Lambda ^2}}{{{N_0}}}}},
\label{eq:29func}
\end{align}
In Eq.~\eqref{eq:29func}, ${{I}_{\partial }}\left( \hbar  \right)=\left( 1/\pi  \right)\int_{0}^{\pi }{{{e}^{\hbar \cos \theta }}\cos \left( \partial \theta  \right)}d\theta $ represents the ${{\partial }^{th}}$ order modified Bessel function of the first kind \cite{Gradshteyn2007Table}, and ${{S}_{\Lambda }}=\sqrt{\frac{\alpha {{E}_{s}}}{{{f}_{num}}{{g}_{num}}}}$. Substituting Eq.~(\ref{eq:29func}) into Eq.~(\ref{eq:28func}) yields
\begin{align}
  {{P}_{ie,\rm{I}|\alpha }} =&\int_{0}^{\infty }{\left[ 1-{{\left[ 1-\exp \left( -\frac{\beta _{\Lambda }^{2}}{{{N}_{0}}} \right) \right]}^{{{N}_{ac}}-{{f}_{num}}}} \right]} \nonumber\\
 &\times \frac{2{{\beta }_{\Lambda }}}{{{N}_{0}}}{{I}_{0}}\left( \frac{2{{S}_{\Lambda }}{{\beta }_{\Lambda }}}{{{N}_{0}}} \right){{e}^{-\frac{\beta _{\Lambda }^{2}+S_{\Lambda }^{2}}{{{N}_{0}}}}}d{{\beta }_{\Lambda }} \nonumber\\
 =&\sum\limits_{q=1}^{{{N}_{g}}-{{f}_{num}}}{{{\left( -1 \right)}^{q}}\left( \begin{matrix}
   {{N}_{ac}}-{{f}_{num}}  \\
   q  \\
\end{matrix} \right)}{{e}^{-\frac{qS_{\Lambda }^{2}}{\left( q+1 \right){{N}_{0}}}}} \nonumber\\
 & \times \int_{0}^{\infty }{\frac{2{{B}_{\Lambda }}}{{{N}_{0}}}{{I}_{0}}\left( \frac{2{{S}_{\Lambda }}{{\beta }_{\Lambda }}}{{{N}_{0}}} \right){{e}^{-\frac{\left( q+1 \right)\beta _{\Lambda }^{2}+\frac{S_{\Lambda }^{2}}{q+1}}{{{N}_{0}}}}}d{{\beta }_{\Lambda }}.}
 \label{eq:30func}
\end{align}
Let ${{{S}'}_{\Lambda }}=\frac{{{S}_{\Lambda }}}{\sqrt{q+1}}$ and ${{{\beta }'}_{\Lambda }}={{\beta }_{\Lambda }}\sqrt{q+1}$ , Eq.~(\ref{eq:30func}) can be rewritten as
\begin{align}
  & {{P}_{ie,\rm{I}|\alpha }}=\sum\limits_{q=1}^{{{N}_{g}}-{{f}_{num}}}{{{\left( -1 \right)}^{q}}\left( \begin{matrix}
   {{N}_{ac}}-{{f}_{num}}  \nonumber\\
   q  \nonumber\\
\end{matrix} \right)}{{e}^{-\frac{qS_{\Lambda }^{2}}{\left( q+1 \right){{N}_{0}}}}} \nonumber\\
 & \times \frac{1}{q+1}\int_{0}^{\infty }{\frac{2{{{{\beta }'}}_{\Lambda }}}{{{N}_{0}}}{{I}_{0}}\left( \frac{2{{{{S}'}}_{\Lambda }}{{{{\beta }'}}_{\Lambda }}}{{{N}_{0}}} \right){{e}^{-\frac{{{\beta }'_{\Lambda }}^{2}+{{S}'_{\Lambda }}^{2}}{{{N}_{0}}}}}d{{{{\beta }'}}_{\Lambda }}} \nonumber\\
 & =\sum\limits_{q=1}^{{{N}_{g}}-{{f}_{num}}}{\frac{{{\left( -1 \right)}^{q}}}{q+1}\left( \begin{matrix}
   {{N}_{ac}}-{{f}_{num}}  \\
   q  \\
\end{matrix} \right)}{{e}^{-\frac{q}{\left( q+1 \right){{f}_{num}}{{g}_{num}}}\cdot \frac{\alpha {{E}_{s}}}{{{N}_{0}}}}}.
\label{eq:31func}
\end{align}
Similar to the derivation of ${P_{ie,{\rm{I}}|\alpha }}$, one can obtain
\begin{align}
{P_{ie,{\rm{II}|}\alpha }}\! =\!\! \sum\limits_{q = 1}^{{N_{ac}} \!- \!{f_{num}}} {\frac{{{{\left( { - 1} \right)}^q}}}{{q + 1}}\left( {\begin{array}{*{20}{c}}
  {{N_{ac}}\!\! - \!\!{f_{num}}} \\
  q
\end{array}} \right)} {e^{ - \frac{q}{{\left( {q + 1} \right){f_{num}}{N_{gs}}}} \cdot \frac{{\alpha {E_s}}}{{{N_0}}}}}.
\label{eq:32func}
\end{align}

\subsubsection{Derivation of ${P_{gie,{\rm{II}|\alpha}}}$}
 According to Eq.~(\ref{eq:13func}), ${P_{gie,{\rm{II}}|\alpha}}$ is expressed as
\begin{align}
{P_{gie,{\rm{II}}|\alpha}} = \Pr \left[ {\mathop {\max }\limits_{\varphi  = 0,1,\ldots,{g_{ac}},\varphi  \ne {g_{{\rm{II}}}}\left( {{\xi}} \right)} \left( {{g_{en,\varphi }}} \right) > {g_{en,{g_{{\rm{II}}}}\left( {{\xi}} \right)}}} \right].
\label{eq:35func}
\end{align}
Conditioned on $\alpha $, ${g_{en,{g_{{\rm{II}}}}\left( {{\xi}} \right)}}$ follows a noncentral chi-square distribution with a noncentrality parameter $S_\lambda ^2 = \frac{{\alpha {E_s}}}{{{N_{gs}}}}$ and $2{N_{ac}}$ degrees of freedom. For notational simplicity, we denote ${g_l} = {g_{en,{g_{{\rm{II}}}}\left( {{\xi}} \right)}}$., whose PDF is given by
\begin{align}
{f_{{g_l}}}\left( {{g_l}} \right) = \frac{1}{{{N_0}}}{\left( {\frac{{{g_l}}}{{S_\lambda ^2}}} \right)^{\frac{{{N_{ac}} - 1}}{2}}}{e^{ - \frac{{S_\lambda ^2 + {g_l}}}{{{N_0}}}}}{I_{{N_{ac}} - 1}}\left( {\frac{{2{S_\lambda }\sqrt {{g_l}} }}{{{N_0}}}} \right).
\label{eq:36func}
\end{align}
Let ${u_n} = {\max _{\varphi  = 0,1,\ldots,{g_{ac}},\varphi  \ne {g_{{\rm{II}}}}\left( {{\xi}} \right)}}\left( {{g_{en,\varphi }}} \right)$ depict  the maximum value of the ${g_{ac}} - {N_{gs}}$ chi-square random variables. The CDF of ${u_n}$ can be formulated as
\begin{align}
{F_{{u_n}}}\left( {{u_n}} \right) = {\left[ {1 - {e^{ - \frac{{{u_n}}}{{{N_0}}}}}\sum\limits_{\kappa  = 0}^{{N_{ac}} - 1} {\frac{1}{{\kappa !}}} {{\left( {\frac{{{u_n}}}{{{N_0}}}} \right)}^\kappa }} \right]^{{g_{ac}} - {N_{gs}}}}.
\label{eq:37func}
\end{align}
According to Eqs.~(\ref{eq:36func}) and (\ref{eq:37func}), Eq.~(\ref{eq:35func}) is obtained as
\begin{align}
  &{P_{gie,{\rm{II}}|\alpha }}\nonumber\\
  &= \!\!\int_0^\infty\!\!  {\left[ {1\! \!-\!\! {{\left[ {1 \!\!- \!\!{e^{ - \frac{{{g_l}}}{{{N_0}}}}}\sum\limits_{\kappa  = 0}^{{N_{ac}} - 1}\!\! {\frac{1}{{\kappa !}}} {{\left( {\frac{{{g_l}}}{{{N_0}}}} \right)}^\kappa }} \right]}^{{g_{ac}} - {N_{gs}}}}} \right]}\!\!  \times \!\! {f_{{g_l}}}\left( {{g_l}} \right)d{g_l} \hfill \nonumber\\
  & = \int_0^\infty  {{f_{{g_l}}}\left( {{g_l}} \right)} \left[ {\sum\limits_{{q_1} = 0}^{{g_{ac}} - {N_{gs}}} {\left( {\begin{array}{*{20}{c}}
  {{g_{ac}} - {N_{gs}}} \\
  {{q_1}}
\end{array}} \right){{\left( { - 1} \right)}^{{q_1} + 1}}} } \right. \hfill \nonumber\\
&   \ \ \ \ \ \times \left. {{{\left( {{e^{ - \frac{{{g_l}}}{{{N_0}}}}}\sum\limits_{\kappa  = 0}^{{N_{ac}} - 1} {\frac{1}{{\kappa !}}{{\left( {\frac{{{g_l}}}{{{N_0}}}} \right)}^\kappa }} } \right)}^{{q_1}}}} \right]d{g_l}. \hfill
\label{eq:38func}
\end{align}
Utilizing \cite[Eq.(0.314)]{Gradshteyn2007Table}, one has
\begin{align}
{\left( {{e^{ - \frac{{{g_l}}}{{{N_0}}}}}\sum\limits_{\kappa  = 0}^{{N_{ac}} - 1} {\frac{1}{{\kappa !}}{{\left( {\frac{{{g_l}}}{{{N_0}}}} \right)}^\kappa }} } \right)^{{q_1}}}\!\! =\!\! {e^{ - \frac{{{q_1}{g_l}}}{{{N_0}}}}}\sum\limits_{\kappa  = 0}^{{q_1}\left( {{N_{ac}} - 1} \right)}\!\! {{\sigma _{\kappa ,{q_1}}}{{\left( {\frac{{{g_l}}}{{{N_0}}}} \right)}^\kappa }},
\label{eq:39func}
\end{align}
where the coefficients ${\sigma _{\kappa ,{q_1}}}$ can be expressed as
\begin{align}
{\sigma _{\kappa ,{q_1}}} = \left\{ \begin{gathered}
  1\ \ \ \ \ \ \ \ \ \ \ \ \ \ \ \ \ \ \ \ \ \ \ \ \ \ \ \ \ \ \ \ \ \ \ \ \ \kappa  = 0 \hfill \\
  \frac{1}{\kappa }\sum\limits_{\ell  = 1}^\kappa  {\left( {\ell {q_1} - \kappa  + \ell } \right){\varpi _\ell }{\sigma _{\kappa  - \ell ,{q_1}}}{\text{  }}\ \kappa  \geqslant 1}  \hfill \\
\end{gathered}  \right.,
\label{eq:40func}
\end{align}
and
\begin{align}
{\varpi _\ell } = \left\{ \begin{gathered}
  \frac{1}{{\ell !}}\ \ \ \ \ \ \ \ell  \leqslant {N_{ac}} - 1 \hfill \\
  0{\text{       \ \ \ \ \ \ \ \ \ Otherwise}} \hfill \\
\end{gathered}  \right.
\label{eq:41func}.
\end{align}
Substituting Eq.~(\ref{eq:39func}) and ${g_l} = {N_0}v$ into Eq.~(\ref{eq:38func}), one can further obtain
\begin{align}
  {P_{gie,{\rm{II}}|\alpha }} =& \sum\limits_{{q_1} = 1}^{{g_{ac}} - {N_{gs}}} {\sum\limits_{\kappa  = 0}^{{q_1}\left( {{N_{ac}} - 1} \right)} {{\sigma _{\kappa ,{q_1}}}\left( {\begin{array}{*{20}{c}}
  {{g_{ac}} - {N_{gs}}} \nonumber\\
  {{q_1}}
\end{array}} \right)} } {\left( { - 1} \right)^{{q_1} + 1}} \hfill \nonumber\\
   &\times \frac{{{e^{ - \frac{{\alpha {E_s}/{N_0}}}{{{N_{gs}}}}}}}}{{{{\left( {\frac{{\alpha {E_s}/{N_0}}}{{{N_{gs}}}}} \right)}^{\frac{{{N_{ac}} - 1}}{2}}}}} \times \int_0^\infty  {{e^{ - v\left( {{q_1} + 1} \right)}}{v^{\left[ {\kappa  + \frac{{\left( {{N_{ac}} - 1} \right)}}{2}} \right]}}}  \hfill \nonumber\\
   &\times {I_{{N_{ac}} - 1}}\left( {2\sqrt {\frac{{v\alpha {E_s}}}{{{N_{gs}}{N_0}}}} } \right)dv. \hfill
   \label{eq:42func}
\end{align}
Utilizing \cite[Eqs.(6.643, 9.215, 9.220)]{Gradshteyn2007Table}, Eq.~(\ref{eq:42func}) can be calculated as
\begin{align}
  &{P_{gie,{\rm{II}}|\alpha }} = \sum\limits_{{q_1} = 1}^{{g_{ac}} - {N_{gs}}} {\sum\limits_{\kappa  = 0}^{{q_1}\left( {{N_{ac}} - 1} \right)} {{\sigma _{\kappa ,{q_1}}}\left( {\begin{array}{*{20}{c}}
  {{g_{ac}} - {N_{gs}}} \nonumber\\
  {{q_1}}
\end{array}} \right)} } {\left( { - 1} \right)^{{q_1} + 1}} \hfill \nonumber\\
  & \times\!\! \frac{{\Gamma \left( {\kappa  + {N_{ac}}} \right){e^{ - \left( {\alpha {E_s}/{N_0}} \right)}}}}{{{{\left( {1 + {q_1}} \right)}^{\left( {\kappa  + {N_{ac}}} \right)}}\Gamma \left( {{N_{ac}}} \right)}}{}_1{F_1}\!\!\left(\!\! {\kappa  + {N_{ac}};{N_{ac}};\frac{{\alpha {E_s}/{N_0}}}{{{N_{gs}}\left( {1 + {q_1}} \right)}}} \!\!\right), \hfill
  \label{eq:43func}
\end{align}
where ${}_1{F_1}\left( { \cdot ; \cdot ; \cdot } \right)$ is the confluent hypergeometric function \cite{Gradshteyn2007Table} and $\Gamma \left(  \cdot  \right)$ is the gamma function \cite{Gradshteyn2007Table}.

Therefore, the BER expressions of scheme I and scheme II can be obtained based on the above derivations.
In particular, for scheme I, the BER expression can be calculated by combining Eqs.~(\ref{eq:17func}), (\ref{eq:18func}), (\ref{eq:pepresult}), and (\ref{eq:31func}). For scheme II, Eqs.~ (\ref{eq:21func}), (\ref{eq:22func}), (\ref{eq:pepresult3}), and (\ref{eq:42func}) are first combined to get ${P_{b,II,gi}}$, then Eqs.~ (\ref{eq:23func}), (\ref{eq:24func}), (\ref{eq:pepresult2}), and (\ref{eq:32func}) are combined to get ${P_{b,{\rm{II}},g}}$. Finally, ${P_{b,{\rm{II}},gi}}$ and ${P_{b,{\rm{II}},g}}$ are substituted into Eq.~ (\ref{eq:20func}) to yield the BER expression.
Although the BER expressions of both schemes are in single-integral form, they can be easily evaluated by utilizing
numerical integration.



\subsection{Transmission Throughput}\label{sect:Throughput Analysis}
A salient feature of the proposed FBI-LoRa system is that it can achieve high-data-rate transmissions, the transmission throughput is a significant metric to verify the superiority of our design. 
In a wireless communication system, the throughput is defined as the number of transmitted bits that can be correctly detected by the receiver, which is expressed as \cite{5934344,8036271,7112587}
\begin{align}
{T_R} = \frac{{{F_{pa}}{N_t}\left( {1 - {P_{pa}}} \right)}}{{{T_s}}}.
\label{eq:ta1}
\end{align}
Here, ${F_{pa}}$ is the number of the symbols in each packet, $N_t$ is the number of information bits carried by each symbol, ${T_s} = {F_{pa}} \cdot {2^{SF}} \cdot {T_{chip}}$ is the transmission period per packet, and ${P_{pa}}$ is the packet error rate, given by
\begin{align}
{P_{pa}} = 1 - {\left( {1 - {P_s}} \right)^{{F_{pa}}}},
\label{eq:ta2}
\end{align}
where ${P_s}$ is the SER for a given LoRa system. 
\begin{figure}[tbp]
\center
\vspace{-0.0cm}
\subfigure[\hspace{-0.8cm}]{ \label{fig:subfig:3a}
\includegraphics[width=3.2in,height=2.4in]{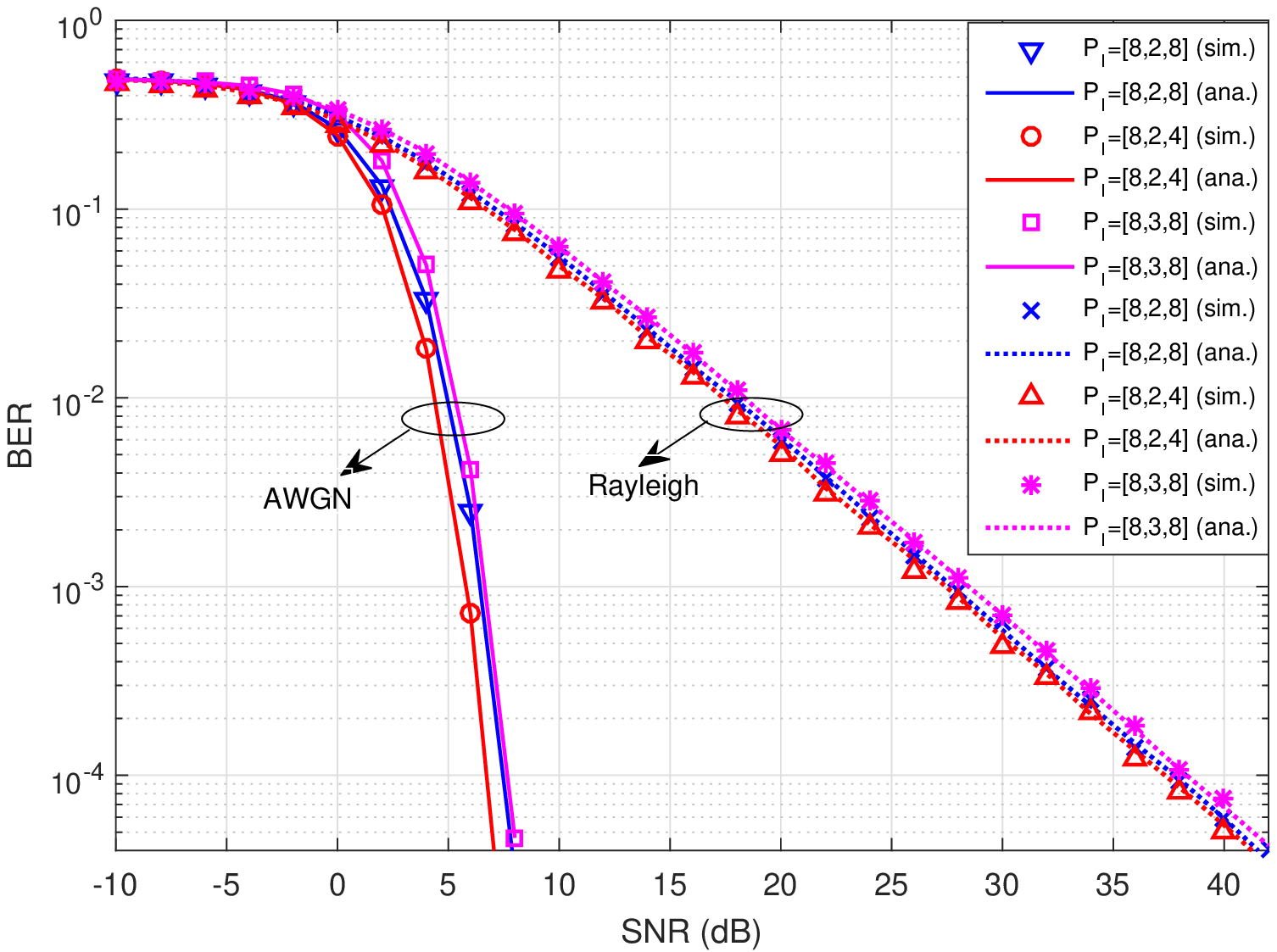}}
\vspace{-0.0cm}
\subfigure[\hspace{-0.8cm}]{ \label{fig:subfig:3b}
\vspace{-0.0cm}
\includegraphics[width=3.2in,height=2.4in]{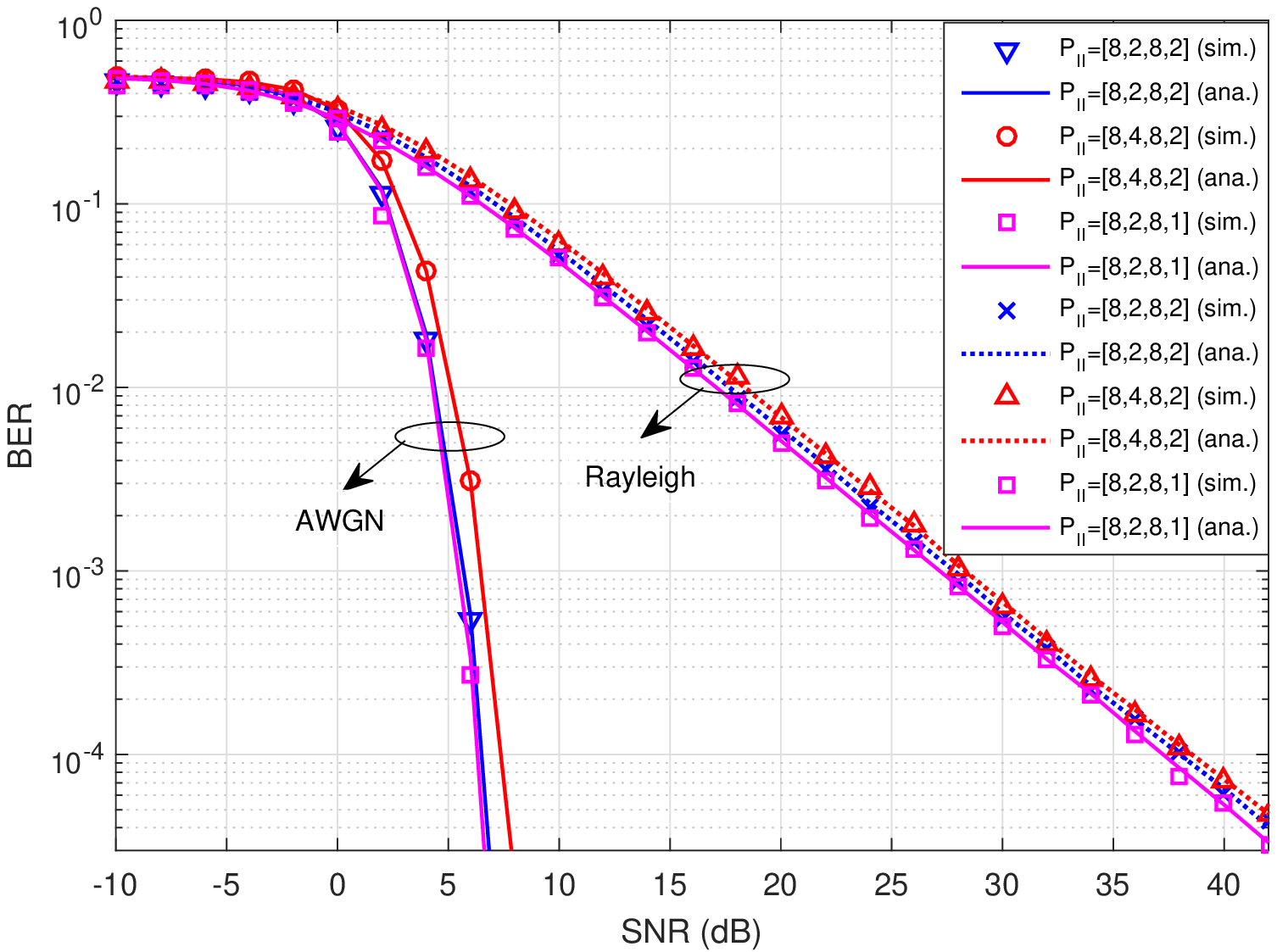}}
\caption{Theoretical and simulated BER results of (a) scheme I and (b) scheme II over AWGN and Rayleigh fading channels.}
\label{fig:fig3}
\vspace{-0mm}
\end{figure}
\begin{figure}[tbp]
\center
\vspace{-0.0cm}
\subfigure[\hspace{-0.8cm}]{ \label{fig:subfig:5a}
\includegraphics[width=3.2in,height=3.2in]{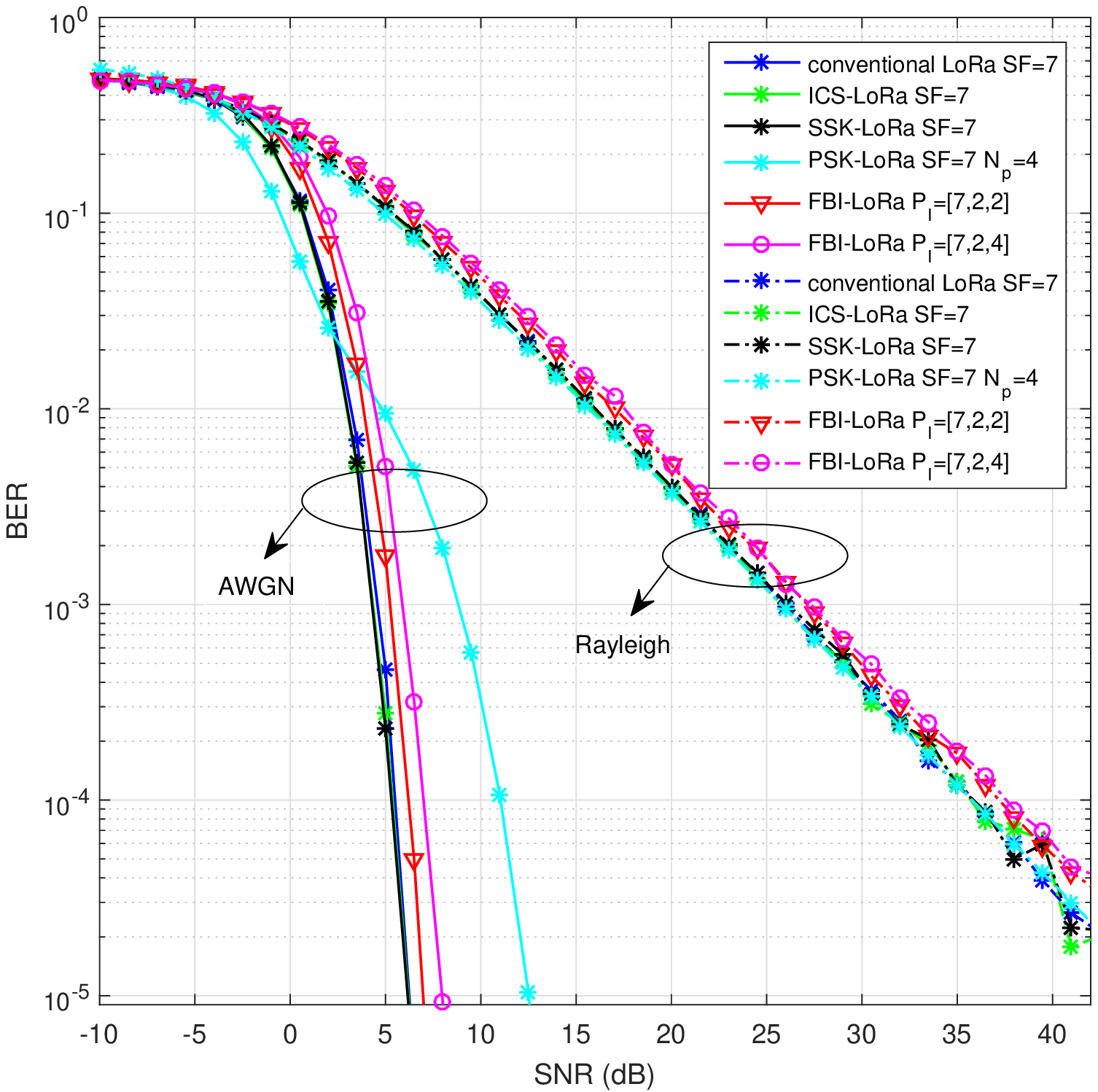}}
\vspace{-0.0cm}
\subfigure[\hspace{-0.8cm}]{ \label{fig:subfig:5b}
\vspace{-0.0cm}
\includegraphics[width=3.2in,height=3.2in]{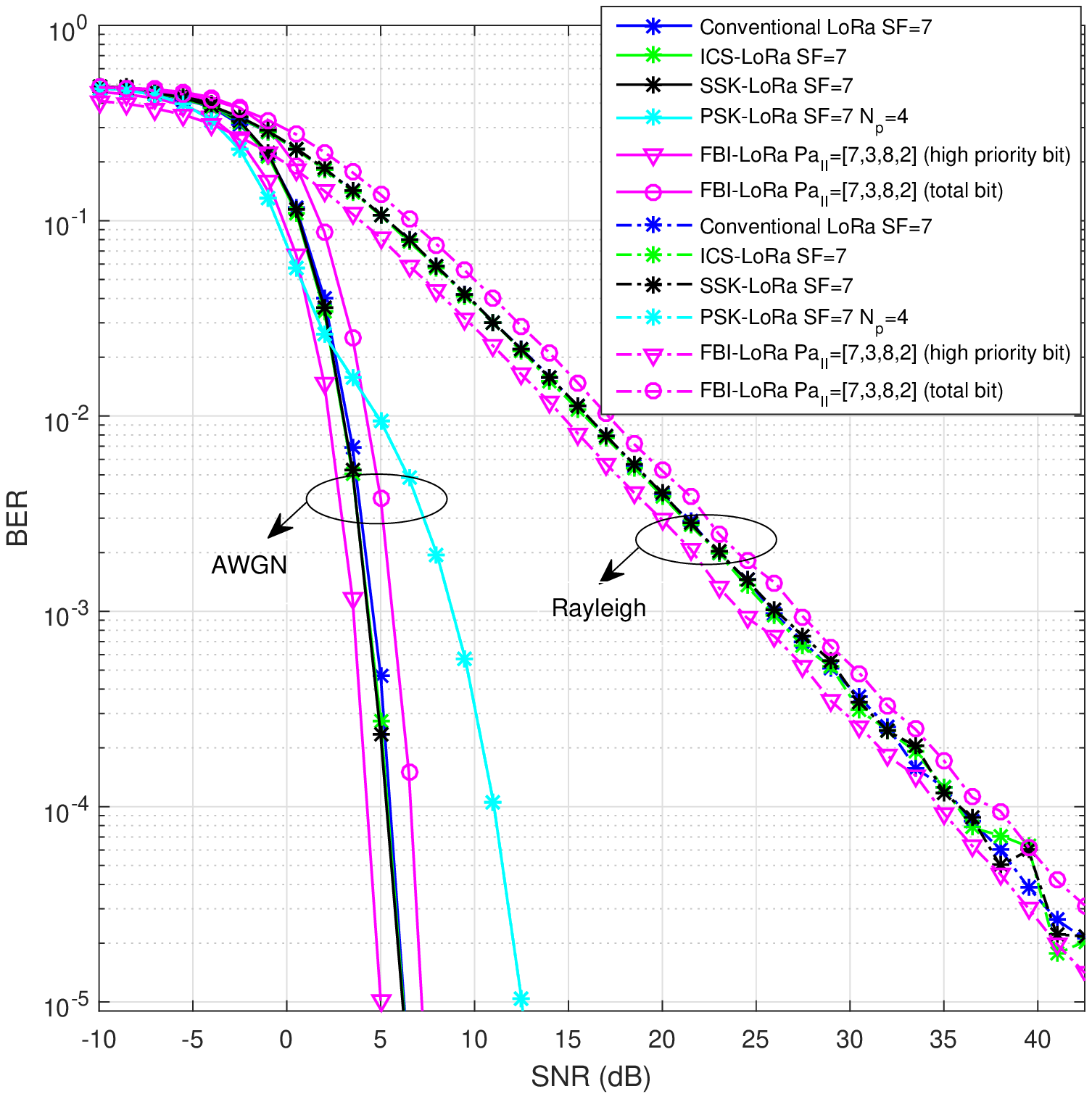}}
\caption{BER performance of (a) scheme I and (b) scheme II compared with that of the conventional LoRa system.}
\label{fig:fig5}
\vspace{-0mm}
\end{figure}
\begin{figure}[tbp]
\center
\vspace{-0.0cm}
\subfigure[\hspace{-0.8cm}]{ \label{fig:subfig:6a}
\includegraphics[width=3.2in,height=2.4in]{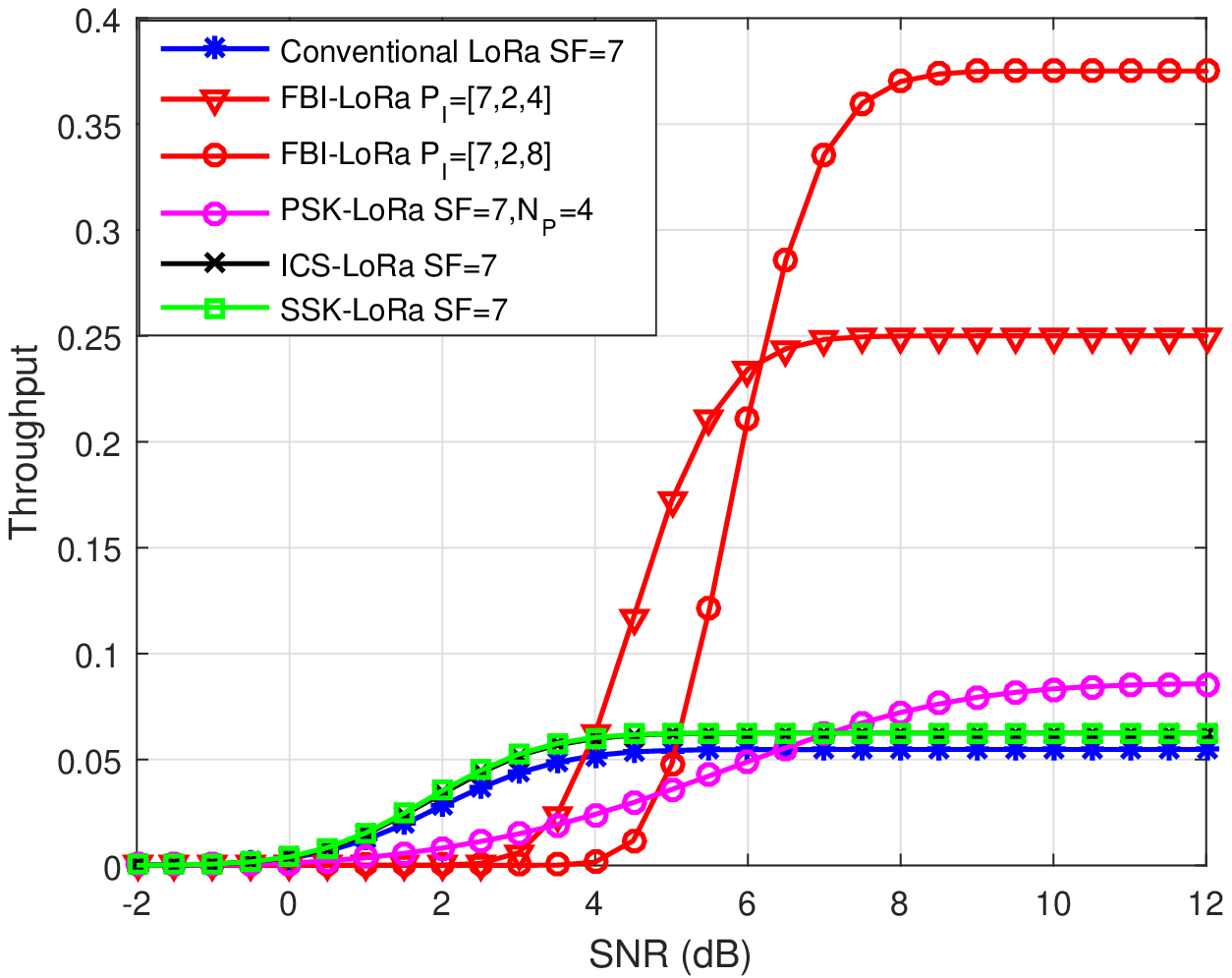}}
\vspace{-0.0cm}
\subfigure[\hspace{-0.8cm}]{ \label{fig:subfig:6b}
\vspace{-0.0cm}
\includegraphics[width=3.2in,height=2.4in]{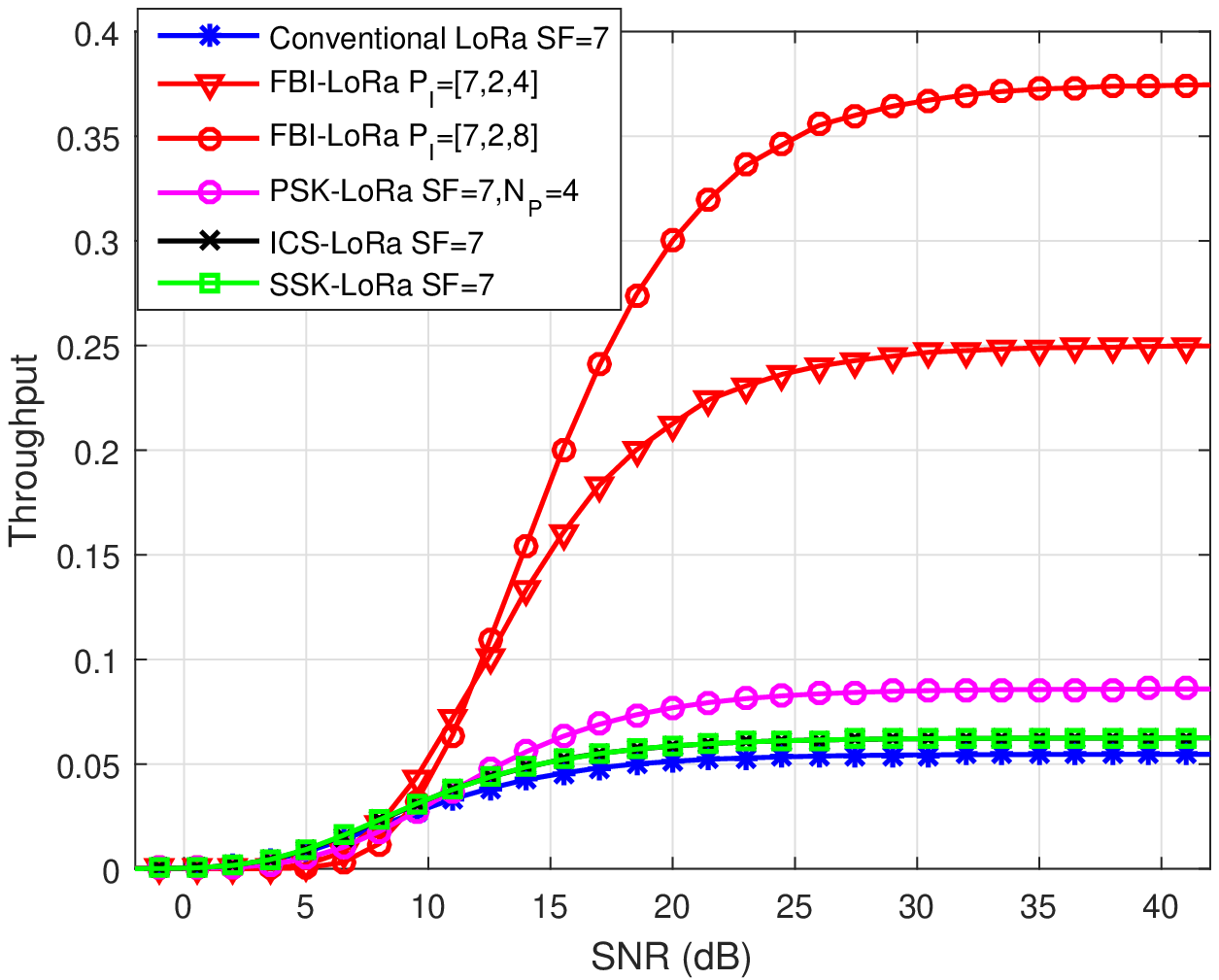}}
\caption{Throughput of scheme I over (a) AWGN and (b) Rayleigh fading channels.}
\label{fig:fig6}
\vspace{-0mm}
\end{figure}

\section{Numerical Results and Discussions}\label{sect:Numerical Results and Discussions}
In this section, we evaluate the BER and the throughput performance of the proposed FBI-LoRa system with scheme I and scheme II over AWGN and Rayleigh fading channels. In the following simulations, the SNR is defined as ${{{E_b}} / {{N_0}}}$, where ${E_b} = {{{E_s}} / {{N_t}}}$ and ${N_t}$ is the number of transmitted bits per symbol. The numerical  and the simulation results in this section are obtained by Mathematica and MATLAB softwares, respectively.

Fig.~\ref{fig:fig3} shows the theoretical and simulated results of the proposed FBI-LoRa system over AWGN and Rayleigh fading channels, where the parameters of scheme I and scheme II are set to ${\bf{Pa}}_{\rm{I}} = \left[ {SF,{f_{num}},{g_{num}}} \right]$ and ${\bf{Pa}}_{{\rm{II}}} = \left[ {SF,{f_{num}},{g_{num}},{N_{gs}}} \right]$, respectively. It can be seen  that the theoretical BER curves agree very well with the simulated ones, for a variety of parameter settings. This substantially verify the accuracy of our theoretical derivation.

In Fig.~\ref{fig:fig5}, we present the BER performance of the proposed FBI-LoRa system and the conventional LoRa system, SSK-LoRa system, ICS-LoRa system, PSK-LoRa system. As shown in Fig.~\ref{fig:subfig:5a} with ${\bf{Pa}}_{\rm{I}} = \left[ {{{7, 2, 2}}} \right]$, compared with the conventional LoRa system, scheme I has performance loss of about $1.4~{\rm dB}$ and $1~{\rm dB}$ at a BER of ${10^{ - 4}}$ over an AWGN channel and a Rayleigh fading channel, respectively. 
Similar observations can be also obtained for other parameter settings.
From Fig.~\ref{fig:subfig:5b}, it can be observed that the BER performance of scheme II is also slightly worse than that of the conventional LoRa system. For instance, scheme II with ${\bf{Pa}}_{{\rm{II}}} = \left[ {7, 3, 8, 2} \right]$ requires about more than $1.2~{\rm dB}$ and $0.8~{\rm dB}$ to achieve a BER of ${10^{ - 4}}$ with respect to the conventional LoRa system. Notably, the ICS-LoRa system and SSK-LoRa system have almost the same BER performance as the conventional LoRa system, and thus the relatively performance between ICS/SSK/convolutiaonl- LoRa system and the propsoed FBI-LoRa system almost remains identical. Furthermore, compared with the PSK-LoRa system, the proposed FBI-LoRa systems with ${\bf{Pa}}_{\rm{I}} = \left[ {{{7, 2, 4}}} \right]$ and ${\bf{Pa}}_{{\rm{II}}} = \left[ {7, 3, 8, 2} \right]$ can achieve around $5.5~{\rm dB}$ and $5~{\rm dB}$ gains at a BER of ${10^{ - 4}}$ over a AWGN channel, respectively. However, the proposed FBI-LoRa system has a performance loss of about $1~{\rm dB}$ at a BER of ${10^{ - 4}}$ compared to the PSK-LoRa (${{n_p} = 4}$) system over a Rayleigh fading channel.

It is noteworthy that, the information bits ${{\bB}_{{\rm{II}},gi}}$ carried by the group index have a higher priority than the information bits ${{\bB}_{{\rm{II}},g}}$ in scheme II because the former enable higher transmitted power. As seen from Fig.~\ref{fig:subfig:5b}, the BER performance of the high-priority bits in scheme II is better than that of the conventional LoRa system. For example, the high-priority bits in scheme II with ${\bf{Pa}}_{{\rm{II}}} = \left[ {7, 3, 8, 2} \right]$ can achieve $1.2~{\rm dB}$ and $1~{\rm dB}$ gains over thoses in the conventional LoRa system at a BER of ${10^{ - 4}}$ over AWGN and Rayleigh fading channels, respectively.
\begin{figure}[tbp]
\center
\vspace{-0.0cm}
\subfigure[\hspace{-0.8cm}]{ \label{fig:subfig:7a}
\includegraphics[width=3.2in,height=2.4in]{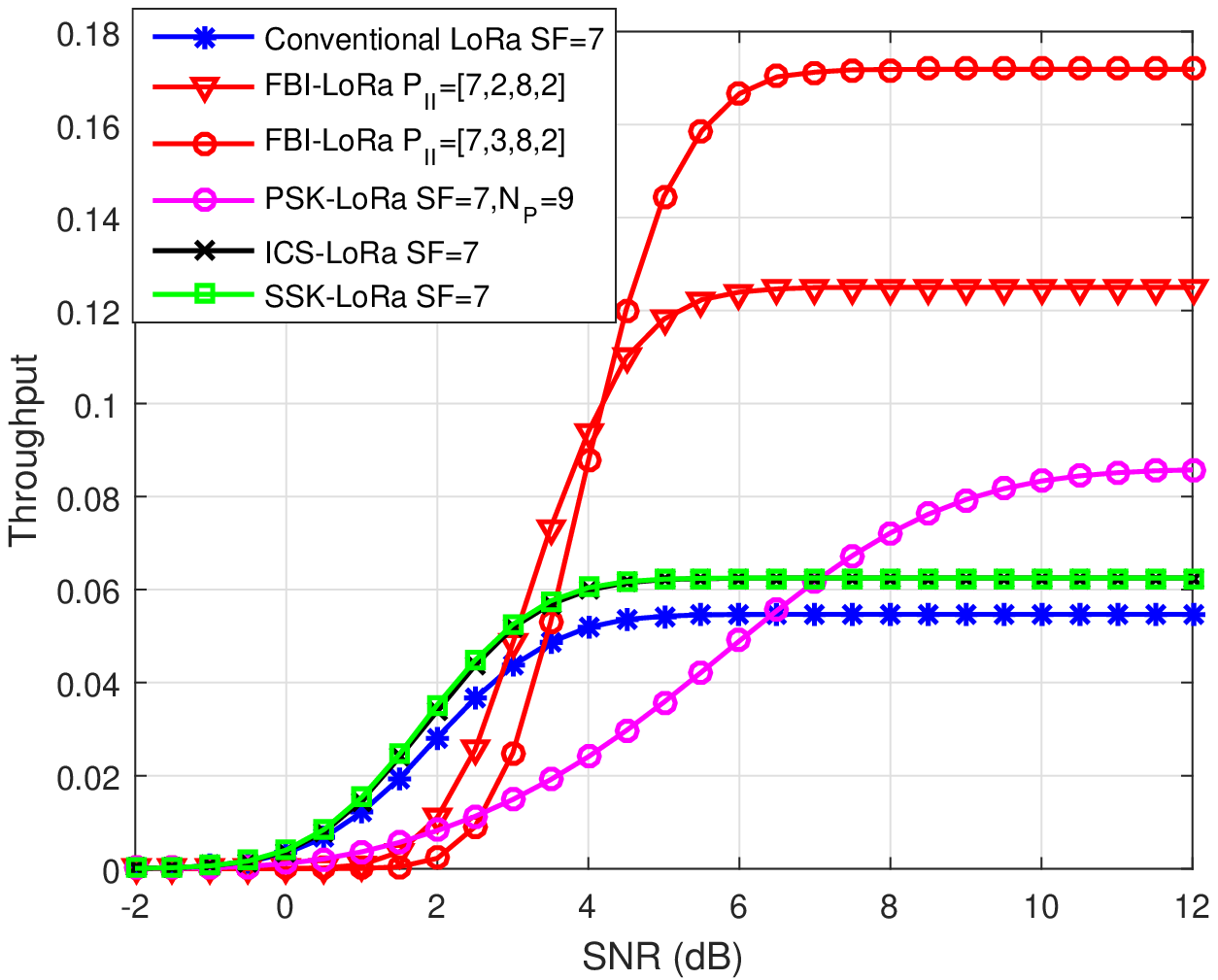}}
\vspace{-0.0cm}
\subfigure[\hspace{-0.8cm}]{ \label{fig:subfig:7b}
\vspace{-0.0cm}
\includegraphics[width=3.2in,height=2.4in]{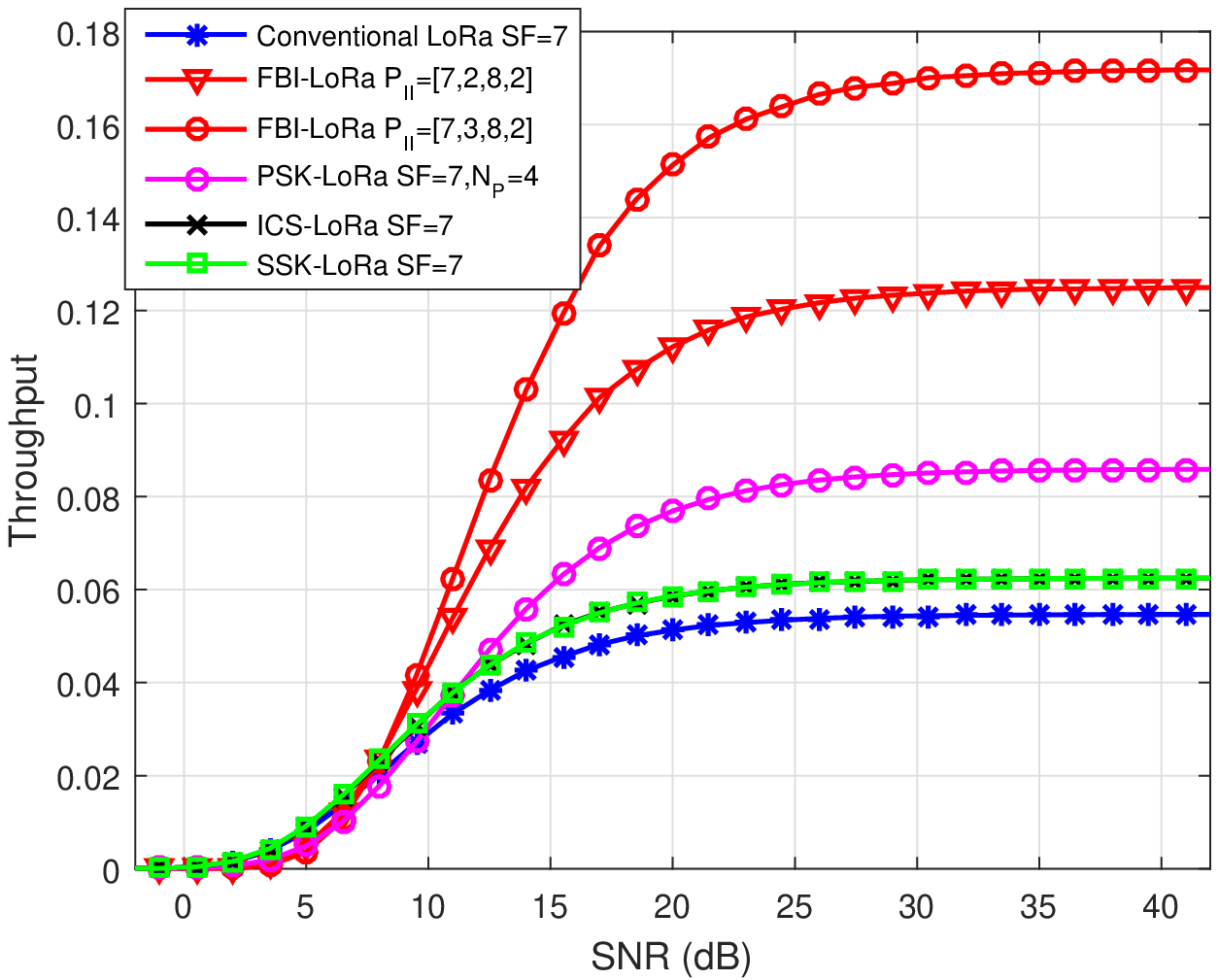}}
\caption{Throughput of scheme II over (a) AWGN and (b) Rayleigh fading channels.}
\label{fig:fig7}
\vspace{-0mm}
\end{figure}
\begin{figure}[tbp]
\center
\vspace{-0.0cm}
\subfigure[\hspace{-0.8cm}]{ \label{fig:subfig:8a}
\includegraphics[width=3.2in,height=2.4in]{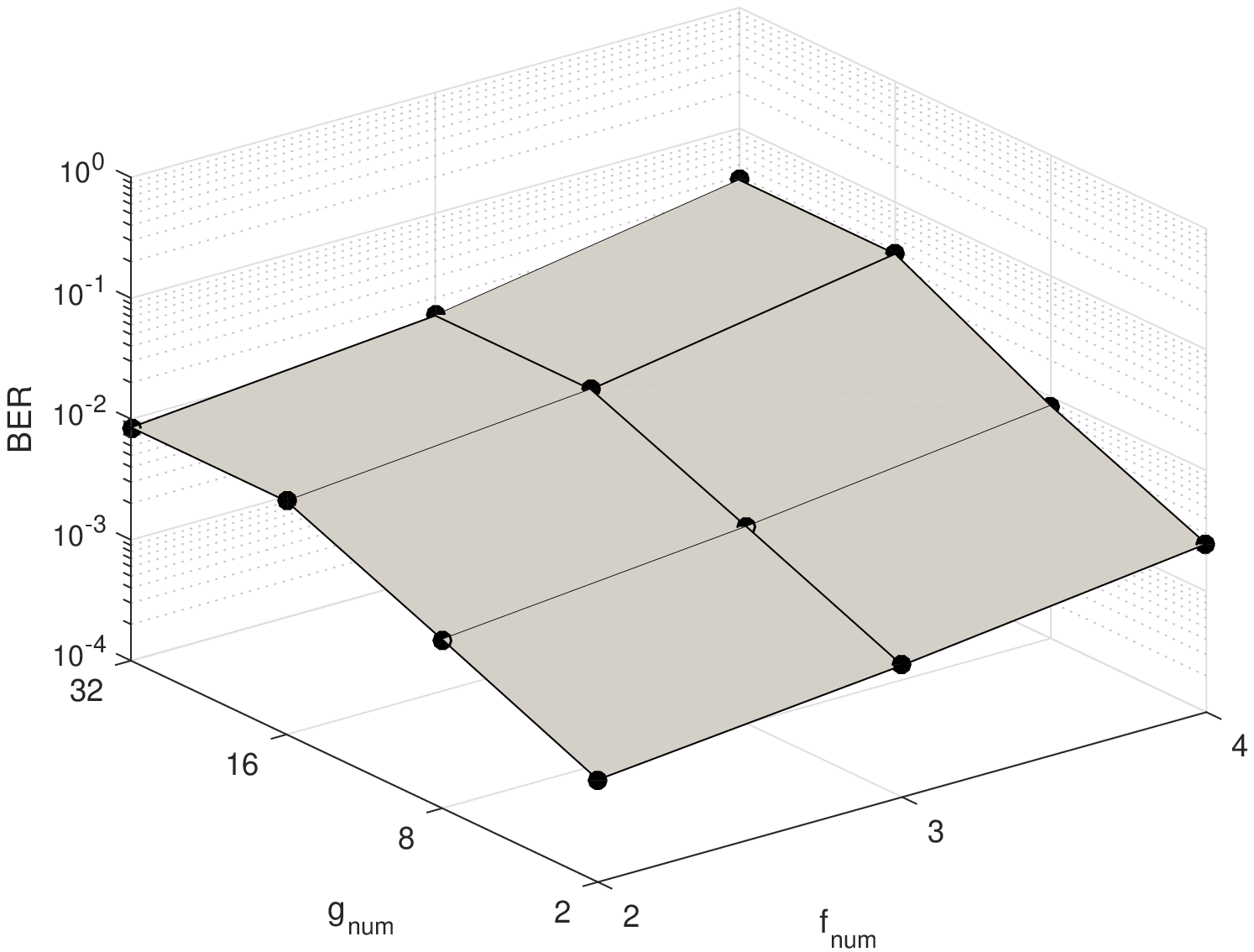}}
\vspace{-0.0cm}
\subfigure[\hspace{-0.8cm}]{ \label{fig:subfig:8b}
\vspace{-0.0cm}
\includegraphics[width=3.2in,height=2.4in]{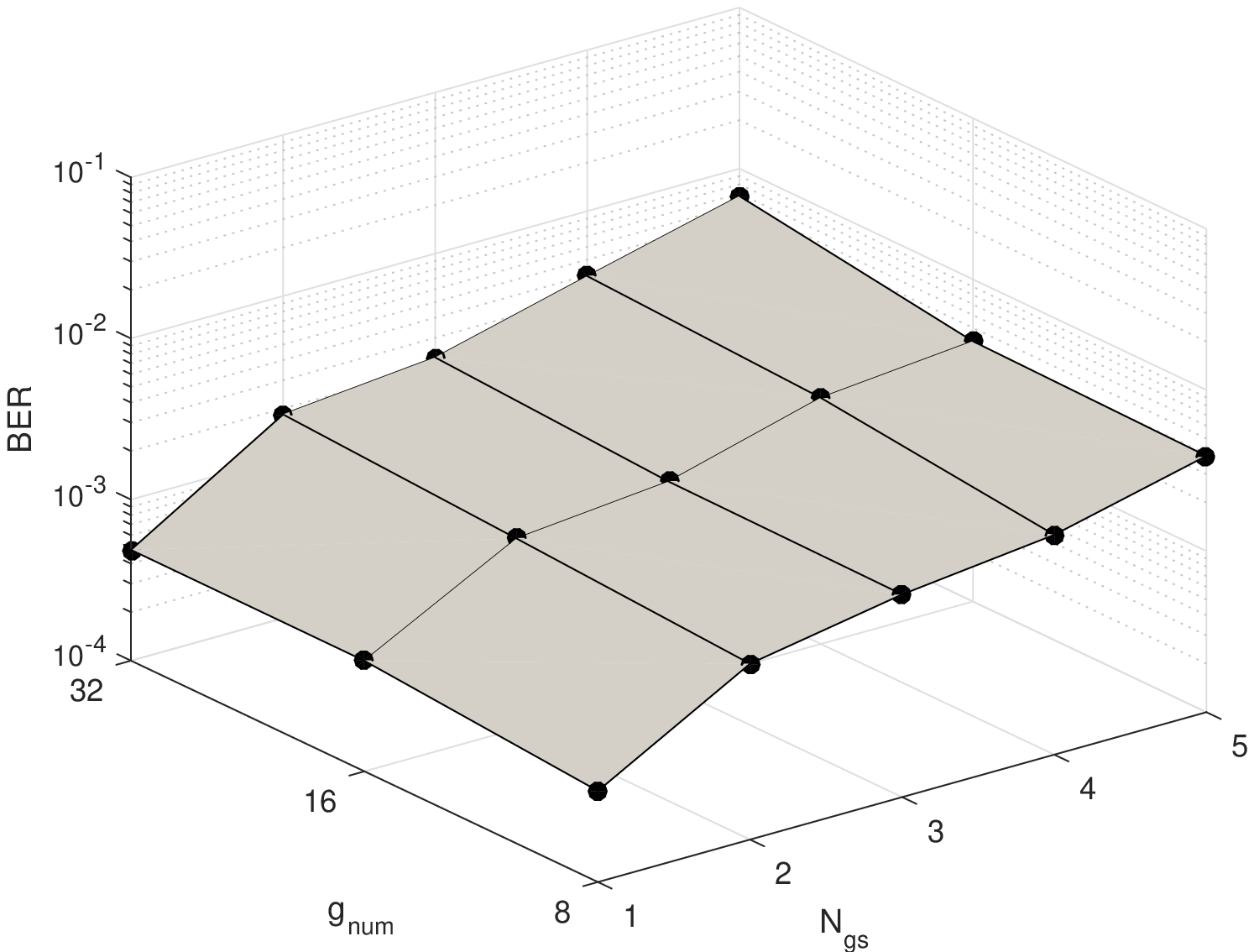}}
\caption{BER performance of (a) scheme I versus $(g_{num}, f_{num})$ and (b) scheme II (${f_{num}} = 2$) versus $(g_{num}, N_{gs})$ over AWGN channels, $SF$ and ${E_b}/{N_0}$ are set to $7$ and $6~{\rm dB}$, respectively.}
\label{fig:fig8}
\vspace{-0mm}
\end{figure}

Fig.~\ref{fig:fig6} and Fig.~\ref{fig:fig7} present the transmission throughput performance for the proposed FBI-LoRa system, the conventional LoRa system, PSK-LoRa system, ICS-LoRa system, and SSK-LoRa system over AWGN and Rayleigh fading channels, where ${F_{pa}}$ is set to $8$. As shown in Fig.~\ref{fig:fig6}, scheme I offers a much higher throughput than other systems in high SNR region. Specifically, the throughput of scheme I with ${\bf{Pa}}_{\rm{I}} = \left[ {{{7}},{{2}},{{4}}} \right]$ are $357\%$, $300\%$, and $191\%$ larger than those of the conventional LoRa, ICS/SSK-LoRa, and PSK-LoRa (${{n_p} = 4}$) systems\footnote{Here, $n_p$ is the number of the information bits carried by the phase\cite{8746470}.}, respectively.
Compared to the PSK-LoRa system, scheme I improves the transmission throughput by $191\%$ over AWGN and Rayleigh fading channels. As expected, similar results appear in Fig.~\ref{fig:fig7}. For instance, the transmission throughput of scheme II with ${\bf{Pa}}_{{\rm{II}}} = \left[ {7,2,8,2} \right]$ are $128.5\%$ and $100\%$ larger than those of the conventional LoRa and ICS/SSK-LoRa systems, respectively. In addition, scheme II has a throughput gain in excess of $45\%$ over the PSK-LoRa (${{n_p} = 4}$) system in both AWGN and Rayleigh fading environments.

Fig.~\ref{fig:fig8} shows the BER performance of scheme I and scheme II under different parameter configurations. It can be seen from Fig.~\ref{fig:subfig:8a} that as either ${f_{num}}$ or ${g_{num}}$ increases (i.e., the number of information bits carried by each symbol also increase), BER performance gradually decreases. Moreover, as shown in Fig.~\ref{fig:subfig:8b}, the BER performance of scheme II gradually decreases as either ${N_{gs}}$ or ${g_{num}}$ increase.

In consequence, the proposed FBI-LoRa system has significant throughput gains compared to the ICS-LoRa system, SSK-LoRa system and PSK-LoRa system. The transmitted symbol of ICS-LoRa system and SSK-LoRa system can carry only one more information bit (i.e., $SF+1$  bits) than the conventional LoRa system regardless of the spreading factor $SF$.
In other words, the transmission-throughput gains of the ICS-LoRa system and the SSK-LoRa system decrease as $SF$ increases compared to the conventional LoRa system. In the proposed FBI-LoRa system, the number of the information bits that carried by a transmitted symbol increases with the spreading factor $SF$ when the parameters ${{f}_{num}}$, ${{g}_{num}}$, and ${{N}_{gs}}$ are fixed.
Considering scheme II with ${\bf{Pa}}_{\rm{II}}=\left[ SF,2,8,2 \right]$.
The FBI-LoRa system achieves $128.5\% $, ${\rm{125\% }}$, ${\rm{122}}{\rm{.2\% }}$, ${\rm{120\% }}$, ${\rm{118}}{\rm{.2\% }}$, and ${\rm{116}}{\rm{.67\% }}$ throughput gains over the conventional LoRa system when $SF$s are set to $7$, $8$, $9$, $10$, $11$, and $12$, respectively. Therefore, as the spreading factor $SF$ increases from $7$ to $12$, the throughput gain reduces only by ${{128.5\%  - 116.67\% } \over {128.5\% }} \approx 9.2\% $. However, for the ICS-LoRa system and the SSK-LoRa system, the throughput gain reduces by ${{14.29\%  - 8.33\% } \over {14.29\% }} \approx {\rm{41}}{\rm{.7\% }}$ \cite{8607020} as $SF$ increases from $7$ to $12$.
For the PSK-LoRa system, although the  throughput can be improved by increasing the parameter ${N_p}$, the BER performance of the PSK-LoRa system dramatically deteriorates simultaneously \cite{8746470}. More importantly, the PSK-LoRa system requires additional channel estimation which increases the hardware complexity.

\begin{figure}[tbp]
\center
\includegraphics[width=3.2in,height=2.4in]{{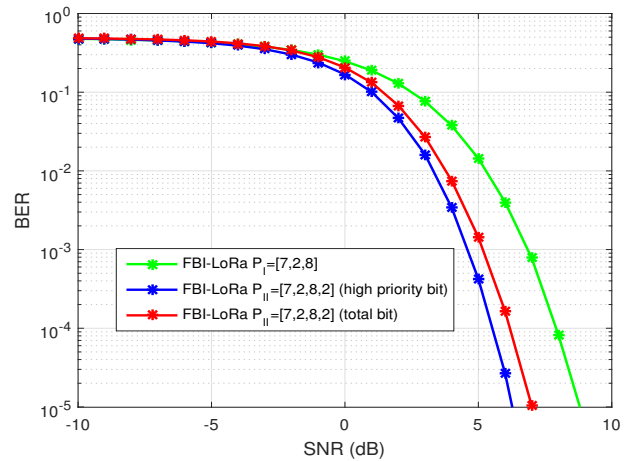}}
\caption{BER performance of scheme I and scheme II conditioned on a fixed parameter setting (i.e., $SF~=~7$, $f_{num}~=~2$, and $g_{num}~=~8$) over an AWGN channel.}
\label{fig:fig9}  
\vspace{-0mm}
\end{figure}

\begin{figure}[tbp]
\center
\includegraphics[width=3.2in,height=2.4in]{{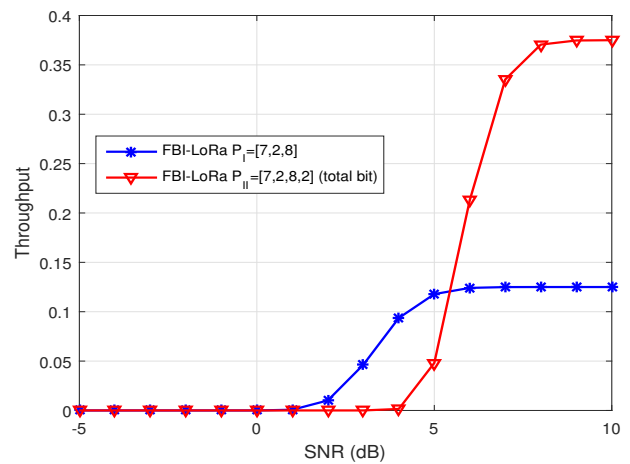}}
\caption{Throughput of scheme I and scheme II conditioned on a fixed parameter setting (i.e., $SF~=~7$, $f_{num}~=~2$, and $g_{num}~=~8$) over an AWGN channel.}
\label{fig:fig10}  
\vspace{-0mm}
\end{figure}

In summary, the FBI-LoRa system can be employed for high-data-rate transmission scenarios, although there exists a tradeoff between throughput and BER performance. Concretely speaking, the FBI-LoRa system can meet different data-rate and error-performance requirements by adjusting parameter setting so as to different practical applications \cite{8390453,SINHA201714,ZORBAS20201,9055222,9065199}.
As seen from Fig.~\ref{fig:fig9} and Fig.~\ref{fig:fig10}, given the fixed parameters $SF$, ${f_{num}}$, and ${g_{num}}$, the BER performance of scheme II is better than that of scheme I, while the throughput performance of scheme I is better than that of scheme II. Consequently,
scheme I and scheme II can serve as excellent transmission solutions for smart home and outdoor scenarios, respectively.
\section{Conclusion}\label{sect:Conclusions}

In this paper, we have studied the design and performance analysis of a new FBI-LoRa system for realizing high-data-rate transmissions in wireless environments. In particular, we have conceived two efficient and practical FBI-aided modulation schemes, referred to as scheme I and scheme II, to achieve this goal. In both schemes, the SFBs of a LoRa signal are divided into several groups so as to keep the low computational complexity of demodulation.
The source information is modulated by the combination of the SFB indices in scheme I, while it is modulated by both the combinations of the group indices and SFB indices. Thanks to the two-dimensional index medium, scheme II is amenable to higher flexibility and better error performance compared with scheme I under a fixed parameter setting.
Furthermore, we have carried out theoretical analyses and simulations to demonstrate the throughput benefit of proposed FBI-LoRa system over AWGN and Rayleigh fading channels, and have discussed the impact of some critical parameters on the system performance.
Results have shown that the proposed FBI-LoRa system significantly outperforms the the conventional LoRa, ICS-LoRa, SSK-LoRa, and PSK-LoRa systems by slightly sacrificing the BER performance.
As a consequence, the proposed FBI-LoRa system can be considered as a promising candidate for practical LPWA applications with high-data-rate requirement.


\end{document}